\documentclass[letterpaper,10pt]{article}

\usepackage{fancyhdr}
\usepackage{epsf}%
\usepackage{hyperref}%
\usepackage{amsmath}%
\usepackage{amsfonts}%
\usepackage{amssymb}%
\usepackage{graphicx}
\usepackage{subfigure}
\usepackage[usenames]{color}
\usepackage[T1]{fontenc} 
\numberwithin{equation}{section}
\def\BK#1{\left\langle #1 \right\rangle}
\def\omit#1{{}}
\def\e#1{\text{e}^{#1}}
\def\tr#1{\text{tr}\left(#1\right)}
\def\d#1{\text{d}{#1}}
\def\pd#1{\partial_{#1}}

\definecolor{red}{rgb}{1,0,0}
	
\definecolor{grey}{rgb}{0.7,0.7,0.7}
	
\newcommand{\Pol}{\operatorname{Pol}\displaylimits}
\newcommand{\Res}{\operatorname{Res}\displaylimits}
\newcommand{\simif}{\operatorname{\sim}\displaylimits}
\newcommand{\eqif}{\operatorname{=}\displaylimits}
\newcommand{\tops}[2]{\texorpdfstring{#1}{#2}}

\def \A {{\cal{A}}}
\def \B {{\cal{B}}}
\def\genus{{\overline{g}}}

\def\O#1{{\cal{O}}\left(#1\right)}
\def \CC {{G_2}}
\newcommand{\cc}[1]{{c_{#1}}} 
\def \L {{\Lambda}}
\def \s {\mathfrak{s}}
\def \w {\omega}
\def \ww {w}

\begin{document}
\hfill SPhT-T08/073\\
\hspace*{\stretch{1}} CRM-3257

\begin{center}
{\Large\bf 

\vspace{12pt}
Topological expansion of the chain of matrices

}

\vspace{\stretch{1}}
{\sl B.\ Eynard}\hspace*{0.05cm}${}^{\dagger}$\hspace*{0.05cm}\footnote{ E-mail: bertrand.eynard@cea.fr }, 
{\sl A.\ Prats Ferrer}\hspace*{0.05cm}${}^{\ddagger}$\hspace*{0.05cm}\footnote{ E-mail: pratsferrer@crm.umontreal.ca }, 

\vspace{\stretch{1}}

{\bf Abstract}
\end{center}

We solve the loop equations to all orders in $1/N^2$, for the Chain of Matrices matrix model (with possibly an external field coupled to the last matrix of the chain). 
We show that the topological expansion of the free energy, is, like for the 1 and 2-matrix model, given by the symplectic invariants of \cite{EOFg}.
As a consequence, we find the double scaling limit explicitly, and we discuss modular properties,  large $N$ asymptotics. We also briefly discuss the limit of an infinite chain of matrices (matrix quantum mechanics).


\vspace{\stretch{2}}

{\small \noindent $\dagger$Institut de Physique Th\'eorique,
CEA, IPhT, F-91191 Gif-sur-Yvette, France,\\
CNRS, URA 2306, F-91191 Gif-sur-Yvette, France.\\
$\ddagger$ Centre de Recherches Mathematiques, Pavillon Andr�-Aisenstadt, Universit� de Montr�al.\\
\indent 2920, Chemin de la tour, Montr�al (Qu�bec) H3T 1J4, CANADA.}

\newpage
\pagestyle{plain}
\setcounter{page}{1}

\section{Introduction}

Since the famous discovery of Brezin, Itzykson, Parisi and Zuber \cite{BIPZ}, it has been known and widely used, that formal matrix integrals are generating functions for the enumeration of discrete surfaces of given topologies (the role of topology was first noticed by 't Hooft \cite{thooft}).
The 1-matrix model is known to count discrete surfaces obtained by gluing polygonal pieces side by side. It is the partition function of random discrete surfaces \cite{BIPZ, ZJDFG}, also called random "maps".

Other matrix models are also partition functions of random discrete surfaces, with additional "colors" on the faces \cite{eynform,ZJDFG}.

In particular, the "2-matrix model" is a partition function of random discrete surfaces, whose polygonal pieces can have two possible colors (or say, two possible spins + or -), and surfaces are counted according to the number of edges separating polygons of different colors, that is polygons with different spins. Thus it counts surfaces with a weight proportional to the exponential of $\sum_{<i,j>} \sigma_i \sigma_j$ (where the sum is over pairs of neighboring pieces, and $\sigma_i$ is the spin of the piece $i$). In other words this is an Ising model on a random discrete surface \cite{Kazakov}.

The most natural generalization is the "Chain of matrices" matrix model. 
It is the generating function for counting discrete surfaces, where pieces can have a color $i\in [1,\dots,n]$, and where each spin configuration on the surface is weighted by $\prod_{<i,j>} (C^{-1})_{i,j}$ where $C$ is a Toeplitz matrix of the form:
\begin{equation}
C= \begin{pmatrix}
g_2^{(1)}    & -\cc{1,2} &          &            &         0   \cr
-\cc{1,2} & g_2^{(2)}    & -\cc{2,3} &            &            \cr
         & \ddots   & \ddots   & \ddots     &            \cr
         &          & \ddots   & \ddots     & -\cc{n-1,n} \cr
0         &          &          & -\cc{n-1,n} & g_2^{(n)}      \cr
\end{pmatrix}
\end{equation}
The partition function for the chain of matrices is the formal small $T$ expansion of the following matrix integral:
\begin{equation}
Z= \int dM_1\dots dM_n\,\,\, \e{-\frac{N}{T} \tr{ \sum_{i=1}^n V_i(M_i) -\sum_{i=1}^{n-1}\cc{i,i+1} M_i M_{i+1}}}
\end{equation}
where $V'_i(0)=0$ and $V''_i(0)=g_2^{(i)}$:
\begin{equation}
V_i(x)= \frac{g_2^{(i)}}{2} x^2 + \sum_{k=3}^{d_i+1} \frac{g_{k}^{(i)}}{k} x^k
\label{eq:pot}
\end{equation}
It is a formal series in $T$, such that
\begin{equation}
\ln{Z} = \sum_{g=0}^\infty \left(\frac{N}{T}\right)^{2-2g}\, F_g
\end{equation}
where:
\begin{equation}
F_g = \sum_v T^v\,\,\sum_{S\in {\mathbb{M}}_g(v)}\,\, \frac{1}{\#{\rm Aut}(S)}\,\, \prod_{i,k} (-g_k^{(i)})^{n_{i,k}(S)}\,\,\,\, \prod_{<i,j>} \left((C^{-1})_{i,j}\right)^{n_{{\rm edges}<i,j>}(S)}
\end{equation}
where ${\mathbb{M}}_g(v)$ is the set of all connected orientable discrete surfaces of genus $g$ with $v$ vertices, with $n_{i,k}$ polygonal pieces of size $k$ (i.e. $k-$angles) of color $i$, and $n_{{\rm edges}<i,j>}$ edges separating colors $i$ and $j$, and $\#$Aut is the number of automorphisms of the surface. 
Notice that for fixed $g$ and $v$, ${\mathbb{M}}_g(v)$ is a finite set, and therefore $F_g$ is indeed a formal series in $T$.

One may also be interested in discrete surfaces with $m$ marked faces, whose generating function is given by:
\begin{equation}
\BK{\tr{\frac{1}{x_1-M_{i_1}}}\dots \tr{\frac{1}{x_m-M_{i_m}}}}_c = \sum_{g=0}^\infty \left(\frac{N}{ T}\right)^{2-2g-m}\, W_{i_1,\dots,i_m}^{(g)}(x_1,\dots,x_m)
\end{equation}
\begin{equation}
W_{i_1,\dots,i_m}^{(g)}(x_1,\dots,x_m) = \sum_{S\in {\mathbb{M}}_{g,i_1,\dots,i_m}}\,\, \frac{T^{\#{\rm vertices}(S)}}{\#{\rm Aut}(S)}\,\, \frac{\prod_{i,k} (-g_k^{(i)})^{n_{i,k}(S)}}
{\prod_{k=1}^m x_k^{l_{i_k}(S)+1} }\,\,\,\, 
\prod_{<i,j>} \left((\CC^{-1})_{i,j}\right)^{n_{{\rm edges}<i,j>}(S)}
\end{equation}
where ${\mathbb{M}}_{g,i_1,\dots,i_m}$ is the set of all connected discrete surfaces of genus $g$, with $n_{i,k}$ $k-$angles of color $i$, and $n_{{\rm edges}<i,j>}$ edges separating colors $i$ and $j$, and
with $m$ marked faces (and with one marked edge on each marked face), of respective perimeters $l_1,\dots,l_m$, and colors $i_1,\dots, i_m$. Again, for fixed $m$ and $g$, there are finitely many such surfaces with a given number of vertices, and the sum is a formal power series in $T$.
Notice that if there is only one marked face $m=1$, i.e. one marked edge, we have a rooted map, and $\#{\rm Aut}(S)=1$.

\bigskip

Recently, the computation of the $F_g$'s and $W^{(g)}$'s was completed for the 1-matrix model ($n=1$) in \cite{eyn1loop, eynch}, and 2-matrix model ($n=2$) \cite{ChEyOr, eoloop2mat, EOFg}, and our goal is to extend the method of \cite{EOFg} to the chain of matrices of arbitrary (but finite) length $n\geq 1$.

\bigskip
In fact, the method of \cite{EOFg} allows to find the solution for a generalization of the chain of matrices, where in addition, the last matrix is coupled to a fixed matrix $M_{n+1}$, called external field.
Matrix models with external fields also have some combinatorial interpretations, and have been studied for various applications.
The most famous is the Kontsevich integral, which is the generating function for intersection numbers \cite{kontsevitch, eynMgnkappa, EOFg}.

\medskip

Here, we solve this more general model.

\medskip

Multimatrix model also play an important role in quantum gravity and string theory, where they play the role of a regularized discrete space-time.
The 1-matrix model, counts discrete surfaces without color, and is a model for quantum gravity without matter, whereas the chain of matrices counts discrete surfaces with $n$ colors, and is interpreted as a model of quantum gravity with some matter field \cite{Kazakovloop, davidRMT, ambjornrmt, KazakovGQ, KazakovGQbis, ZJDFG}, namely a matter which can have $n$ possible states.
More recently, matrix models have played a role in topological string theory \cite{DW}.

\vspace{1cm}
\noindent{\bf Outline of the article:}

\begin{itemize}
	\item {In section \ref{Sec:Not} we introduce all the definitions and notations necessary for the derivation of the loop equations. These are quite clearly inspired by the work on \cite{eynchainloopeq} where the loop equations were already found in a slightly less compact way.}
	\item {In section \ref{Sec:MLE} we derive the master loop equation that will allow us to solve the model. We also consider the $\frac{1}{N^2}$ expansion here and find the spectral curve for this model.}
	\item {In section \ref{Sec:AC} we overview all the important algebraic geometry tools and the algebraic curve properties that are relevant for us.}
	\item {In section \ref{Sec:SMLE} we apply the same techniques of \cite{ChEyOr} to prove uniqueness of the solution and to find the actual solution for the correlators of the first matrix $M_1$ of the chain.}
	\item {In section \ref{Moduli} we find the variation of the curve, and all the correlation functions, in terms of the moduli of the chain of matrices. This leads us to an expression for the whole topological expansion of the free energy for the chain of matrices.}
        \item {In section \ref{Sec:OC}, we study some corollaries of the properties of the symplectic invariants of \cite{EOFg}, in particular we get the double scaling limit, and modular properties.}
	\item {In section \ref{Sec:QMM}, we briefly discuss the "matrix quantum mechanics", i.e. the limit of an infinite chain of matrices.}
	\item {Finally, section \ref{Sec:Concl} is the conclusion.}
\end{itemize}

\section{Notations and Definitions}\label{Sec:Not}

\subsection{The \tops{\emph{formal}}{formal} chain matrix model with external field}

The \emph{formal} chain matrix model with external field, is a formal matrix integral\footnote{A formal integral is defined as the exchange of the integral and the Taylor expansion of the exponential of non-quadratic terms in the potentials, see \cite{eynform}.}, with $n$ matrices of size $N$ with potentials $V_i(M_i)$, arranged in a chain with Itzykson-Zuber like interactions:
\begin{equation}
	Z_{Ch}=\int \prod_{i=1}^n \d{M_i}\,
    \e{-\frac{N}{T}\tr{\sum_{i=1}^{n}\left(V_{i}(M_i)-\cc{i,i+1}M_iM_{i+1}\right)}}
	\label{eq:COM}
\end{equation}
where $M_{n+1}$ is a constant matrix, which we may choose diagonal $M_{n+1}=\Lambda$ ,  with $\s$ different eigenvalues $\lambda_i$ and multiplicities $l_i$ ($\sum_i l_i=N$):
\begin{equation}
    \Lambda=\text{diag}(\underbrace{\lambda_1,\cdots,\lambda_1}_{l_1}
    ,\cdots,\underbrace{\lambda_i,\cdots,\lambda_i}_{l_i}
    ,\cdots,\underbrace{\lambda_\s,\cdots,\lambda_\s}_{l_\s}).
\end{equation}
It reduces to the usual "Chain of Matrices" when $\L=M_{n+1}=0$.

The measures $\d{M_i}=\prod_{j=1}^N \d{M_{jj}^{(i)}}\prod_{j<k}^N \d{\Re(M_{jk}^{(i)})}\d{\Im(M_{jk}^{(i)})}$ are the usual Lebesgue measures for hermitian matrices.
The potentials $V_i(x)$ are polynomials\footnote{Notice that here, in contrast to equation \eqref{eq:pot}, we allow for a linear term in the potential. This is convenient but can be trivially undone by a shifts proportional to the identity.} of degree $d_i+1$, 
\begin{equation}
    V_i(x)=\sum_{k=1}^{d_i+1} \frac{g_k^{(i)}}{k} x^k
\end{equation}
but the same results contained in this paper can clearly be extended to functions $V_i$ whose derivatives $V'_i$ are rational functions.
In general we are interested in formal expectation values of functions of $M_i$ defined by
\begin{equation}
    \BK{f(M_1,\cdots,M_n)}=\frac{1}{Z_{Ch}}\int \prod_{i=1}^n \d{M_i}\,f(M_1,\cdots,M_n)\,
    \e{-\frac{N}{T}\tr{\sum_{i=1}^{n}\left(V_{i}(M_i)-\cc{i,i+1} M_iM_{i+1}\right)}}
\end{equation}
but we will also be interested in the free energy defined as the logarithm of the partition function $Z_{Ch}$.

The $\frac{1}{N^2}$ expansion can be considered when we work with the formal version of this matrix integral. What that means is that we must interpret the integrals as a formal expansion of all the non-gaussian terms in the exponential and perform the integral as a perturbation integral around a minimum of the action 
\begin{equation}
	\tr{\sum_i V_i(M_i) - \sum_i \cc{i,i+1}M_i M_{i+1}}.
\end{equation}
The equations that define a minimum are
\begin{equation}
V'_1(M_1) = \cc{1,2} M_2
\quad , \quad
V'_k(M_k) = \cc{k-1,k}M_{k-1}+\cc{k,k+1}M_{k+1} \quad k\geq 2
\end{equation}
In particular we can choose a minimum such that all $M_k$'s are diagonal $M_k={\rm diag}(\bar{\mu}_1^{(k)},\dots,\bar{\mu}_N^{(k)})$, which satisfy:
\begin{equation}
\begin{split}
    V_1^\prime(\bar{\mu}_i^{(1)})&=\cc{1,2}\bar{\mu}_i^{(2)}\\
    V_k^\prime(\bar{\mu}_i^{(k)})&=\cc{k-1,k}\bar{\mu}_i^{(k-1)}+\cc{k,k+1}\bar{\mu}_i^{(k+1)}
    \,\quad \,k=2,\cdots,n
\end{split}
    \label{eq:FP}
\end{equation}
with $\bar{\mu}_i^{(n+1)}=\mu_i^{(n+1)}=\Lambda_i$.
Note that $\cc{n,n+1}$ can be absorbed into $\Lambda$, so that we will fix it to 1.
These equations have $D=d_1d_2\cdots d_n \s $ solutions.
Choosing which minimum we are going to perturb around, means choosing how many eigenvalues we are going to put on each of the $D$ different solutions. Let us call these $n_1,\cdots,n_D$, with the restriction $\sum_i n_i=N$. 
In the following we are going to refer to $\epsilon_i=T\,\frac{n_i}{N}$ as the filling fractions.

In other words, for each choice of filling fractions $\epsilon=(\epsilon_1,\dots,\epsilon_{D-1})$, we can define a formal integral by perturbation around the corresponding minimum.
Almost by definition, there must exist anti-clockwise contours ${\cal A}_i$, $i=1,\dots, D$, such that
\begin{equation}
- \frac{T}{2i\pi\, N}\oint_{{\cal A}_i}\,\, \BK{\tr{\frac{1}{x-M_1}}}\,\, dx = \epsilon_i = T\,\frac{n_i}{N}
\end{equation}

\subsection{Definitions of correlation functions}

In order to define the good observables of our model, we first need to introduce (like in \cite{eynchainloopeq}) the following polynomials $f_{i,j}(x_1,\dots,x_n)$
\begin{equation}
\begin{split}
f_{i,j}(x_i,\cdots,x_j)&=\prod_{k=i}^j\frac{1}{\cc{k-1,k}}\det\left|\begin{array}{c@{}c@{}c@{}c}
V_i^\prime(x_i)       & -\cc{i,i+1} x_{i+1} &                    & 0                     \\
-\cc{i,i+1}x_i & V_{i+1}^\prime(x_{i+1})   & \ddots             &                       \\
                      & \ddots                    & \ddots             & -\cc{j,j+1} x_j \\
 0                    &                           & -\cc{j,j+1} x_{j-1} & V_j^\prime(x_j)
\end{array}\right|
\quad {\rm if}\, i\leq j\\
&=1\,\, {\rm if}\, i=j+1\\
&=0\,\, {\rm if}\, i>j+1
\end{split}
\end{equation}
They satisfy the recursion relation
\begin{equation}
	\cc{i-1,i}f_{i,j}(x_i,\dots,x_j)=
    V_{i}^\prime(x_i)f_{i+1,j}(x_{i+1},\dots,x_j)
    -\cc{i,i+1}x_{i}x_{i+1}f_{i+2,j}(x_{i+2},\dots,x_j)
	\label{eq:PRR}
\end{equation}
with the initial conditions $f_{k+1,k}=1$, and $f_{k+l,k}=0$ for all $l>1$.
The first polynomials generated by this recursion relation are
\begin{equation}
	\begin{split}
	f_{i,i}(x_i)&=\frac{V_i^\prime(x_i)}{\cc{i-1,i}}\\
	f_{i-1,i}(x_{i-1},x_i)&=\frac{V_{i-1}^\prime(x_{i-1})}{\cc{i-2,i-1}}\frac{V_{i}^\prime(x_{i})}{\cc{i-1,i}}
	-\frac{\cc{i-1,i}}{\cc{i-2,i-1}}x_{i-1}x_i\\
	f_{i-2,i}(x_{i-2},x_{i-1},x_i)&=
	\frac{V_{i-2}^\prime(x_{i-2})}{\cc{i-3,i-2}}
	\frac{V_{i-1}^\prime(x_{i-1})}{\cc{i-2,i-1}}\frac{V_{i}^\prime(x_{i})}{\cc{i-1,i}}\\
	&-\frac{V_{i-2}^\prime(x_{i-2})}{\cc{i-3,i-2}}\frac{\cc{i-1,i}}{\cc{i-2,i-1}}x_{i-1}x_i-
	\frac{\cc{i-2,i-1}}{\cc{i-3,i-2}}x_{i-2}x_{i-1}\frac{V_{i}^\prime(x_{i})}{\cc{i-1,i}}\\
	\end{split}
	\label{eq:PGN}
\end{equation}
Define also the following functions
\begin{equation}
	\ww_{i}(x_i)=\frac{1}{x_i-M_i}\quad,\qquad Q(z)=\frac{1}{\cc{n,n+1}}\frac{S(z)-S(\Lambda)}{z-\Lambda}
	\label{eq:RAS}
\end{equation}
where $S(z)$  is the minimal polynomial of $\Lambda$:
\begin{equation}
S(z)=\prod_{i=1}^\s (z-\lambda_i)
\end{equation}
The loop equations in following sections will be written in terms of the following matrix model observables or correlation functions.
\begin{equation}
\begin{split}
	W_0(x_1)&=\BK{\frac{T}{N}\tr{\ww_1(x_1)}}\\
	P(x_1)&=\Pol_{x_1}f_{1,1}(x_1)W_0(x_1)=\Pol_{x_1}V_1^\prime(x_1)W_0(x_1)\\
	W_i(x_1,x_i,\dots,x_n,z)&=\Pol_{x_{i},\dotsc,x_n} f_{i,n}(x_{i},\dotsc,x_n)
	\BK{\frac{T}{N}\tr{\ww_1(x_1)\ww_{i}(x_{i})\cdots\ww_n(x_n)Q(z)}},\\
	&\qquad \text{for $i=2,\dotsc,n-1$}\\
	W_1(x_1,\dotsc,x_n,z)&=\Pol_{x_{1},\dotsc,x_n} f_{1,n}(x_{1},\dotsc,x_n)
	\BK{\frac{T}{N}\tr{\ww_1(x_1)\cdots\ww_n(x_n)Q(z)}}\\
	&\qquad \rightarrow\textrm{is a polynomial in all variables}\\
	W_n(x_1,z)&=\BK{\frac{T}{N}\tr{\ww_1(x_1)Q(z)}}\\
	W_{0;1}(x_1;x_1^\prime)&=
    \frac{\partial}{\partial V_1(x_1^\prime)}W_0(x_1)
    =\BK{\tr{\ww_1(x_1)}\tr{\ww_1(x_1^\prime)}}_c\\
	W_{i;1}(x_1,x_i,\dots,x_n,z;x_1^\prime)&=\frac{\partial}{\partial V_1(x_1^\prime)}W_i(x_1,x_i,\dots,x_n,z)=\\
	&=\BK{\tr{\ww_1(x_1^\prime)}\Pol_{x_{i},\dotsc,x_n} f_{i,n}(x_{i},\dotsc,x_n)
	\tr{\ww_1(x_1)\ww_{i}(x_{i})\cdots\ww_n(x_n)Q(z)}}_c
\end{split}
	\label{eq:MMO}
\end{equation}
where the symbol ``$\Pol_x{f(x)}$'' represents the polynomial part on the variable $x$ in the vicinity of $\infty$ of the function ``f(x)'', and the loop insertion operator simbol is defined by 
 \begin{equation}
   \frac{\pd{}}{\pd{} V_i(x)}=-\frac{1}{x}\frac{\pd{}}{\pd{}g_0^{(i)}}-\sum_k \frac{k}{x^{k+1}}\frac{\pd{}}{\pd{} g_k^{(i)}}
 \end{equation} 
like in \cite{eynchainloopeq}.
At some point we will write the topological expansion\footnote{The topological expansion of a formal integral, is not a large $N$ expansion, it is a small $T$ expansion, and for each power of $T$, the coefficient is a polynomial in $N^{-2}$. The $W_0^{(h)}(x)$ is merely the formal series in $T$, containing the degree $h$ terms.} of some of these functions, for example
\begin{equation}
    W_0(x_1)=\sum_{h=0}^\infty \left(\frac{T}{N}\right)^{2h}\,W_0^{(h)}(x_1),
\end{equation}
and similarly for other functions. These are all the definitions we need for the derivation of the loop equations

\section{Master Loop Equation}\label{Sec:MLE}

To find the master loop equation (proceeding as in \cite{eynchainloopeq}) we are going to consider the following local changes of variables
\begin{equation}
	\begin{split}
	\delta M_i &=\epsilon \Pol_{x_{i+1},\dotsc,x_n}f_{i+1,n}(x_{i+1},\dotsc,x_n)
	\ww_{i+1}(x_{i+1})\dotsm\ww_{n}(x_n)Q(z)\ww_1(x_1) +O(\epsilon^2),\quad 1\le i<n\\
	\delta M_n &=\epsilon Q(z)\ww_1(x_1)+O(\epsilon^2)
	\end{split}
	\label{eq:COV}
\end{equation}
with $\epsilon$ a small parameter.
Notice that $\delta M_i$ does not contain $M_i$ except for $i=1$. We must then consider $\delta M_1$ separately.

\subsection{Loop Equation for \tops{$\delta M_1$}{dM1}}\label{sSec:LE1}

Consider the change of variables
\begin{equation}
\delta M_1 =\epsilon \Pol_{x_{2},\dotsc,x_n}f_{2,n}(x_{2},\dotsc,x_n)
	\ww_{2}(x_{2})\dotsm\ww_{n}(x_n)Q(z)\ww_1(x_1)+O(\epsilon^2)\\
	\label{eq:CV1}
\end{equation}
The first order variation in $\epsilon$ of the integral \eqref{eq:COM} gives the Schwinger-Dyson equation (called loop equation in the matrix model context):
\begin{multline}
	\BK{\frac{T^2}{N^2}\tr{\ww_1(x_1)}\tr{\ww_1(x_1)
    \Pol_{x_{2},\dotsc,x_n}f_{2,n}(x_{2},\dotsc,x_n)
	\ww_{2}(x_{2})\dotsm\ww_{n}(x_n)Q(z)}}\\
	=\BK{\frac{T}{N}\tr{\ww_1(x_1)\left(V_1^\prime(M_1)-\cc{1,2}M_2\right)
    \Pol_{x_{2},\dotsc,x_n}f_{2,n}(x_{2},\dotsc,x_n)
	\ww_{2}(x_{2})\dotsm\ww_{n}(x_n)Q(z)}}
	\label{eq:LE1a}
\end{multline}
Using \eqref{eq:PRR} we find, after some algebra, the loop equation
\begin{equation}
\begin{split}
	\frac{T^2}{N^2}W_{2;1}&(x_1,x_2,\dotsc,x_n,z;x_1)
    +(\cc{1,2}x_2-V_1^\prime(x_1)+W_0(x_1))W_2(x_1,x_2,\dotsc,x_n,z)=\\
	=&-W_1(x_1,\dotsc,x_n,z)+(V_2^\prime(x_2)-\cc{1,2}x_1)W_3(x_1,x_3,\dotsc,x_n,z)
    -\cc{2,3}x_2W_4(x_1,x_4,\dotsc,x_n,z)\\
	&-\BK{\frac{T}{N}\tr{\ww_1(x_1)\left(V_2^\prime(M_2)-\cc{1,2}M_1\right)\Pol_{x_{3},\dotsc,x_n}
	f_{3,n}(x_{3},\dotsc,x_n)\ww_{3}(x_{3})\dotsm\ww_{n}(x_n)Q(z)}}\\
	&+\BK{\frac{T}{N}\tr{\ww_1(x_1)\cc{2,3}M_2\Pol_{x_{4},\dotsc,x_n}
	f_{4,n}(x_{4},\dotsc,x_n)\ww_{4}(x_{4})\dotsm\ww_{n}(x_n)Q(z)}}
\end{split}
	\label{eq:LE1z}
\end{equation}

\subsection{Loop Equation for \tops{$\delta M_i$}{dMi}}\label{sSec:LE2}
The rest of the loop equations follow the same principle. We will compute the remaining in one shot.
\begin{equation}
    \delta M_i =\epsilon \Pol_{x_{i+1},\dotsc,x_n}f_{i+1,n}(x_{i+1},\dotsc,x_n)
	\ww_{i+1}(x_{i+1})\dotsm\ww_{n}(x_n)Q(z)\ww_1(x_1)+O(\epsilon^2)
	\label{eq:CV2}
\end{equation}
from which the order $\epsilon$ variation of the partition function is
{\small
\begin{equation}
\begin{split}
	&0=
	\BK{\frac{T}{N}\tr{\ww_1(x_1)\left(V_i^\prime(M_i)-\cc{i-1,i}M_{i-1}\right)
	\Pol_{x_{i+1},x_n}f_{i+1,n}(x_{i+1},\dotsc,x_n)
	\ww_{i+1}(x_{i+1})\dotsm\ww_{n}(x_n)Q(z)}}\\
	&-\cc{i,i+1}x_{i+1}W_{i+1}(x_1,x_{i+1},\dotsc,x_{n},z)+
    V_{i+1}^\prime(x_{i+1})W_{i+2}(x_1,x_{i+2},\dotsc,x_n,z)
	-\cc{i,i+1}x_{i+1}W_{i+3}(x_1,x_{i+3},\dotsc,x_{n},z)\\
	&-\BK{\frac{T}{N}\tr{\ww_1(x_1)V_i^\prime(M_i)\Pol_{x_{i+2},\dotsc,x_n}
	f_{i+2,n}(x_{i+2},\dotsc,x_n)\ww_{i+2}(x_{i+2})\dotsm\ww_{n}(x_n)Q(z)}}\\
	&+\BK{\frac{T}{N}\tr{\ww_1(x_1)\cc{i,i+1}M_{i+1}\Pol_{x_{i+3},\dotsc,x_n}
	f_{i+3,n}(x_{i+3},\dotsc,x_n)\ww_{i+3}(x_{i+3})\dotsm\ww_{n}(x_n)Q(z)}}
\end{split}
	\label{eq:LE2a}
\end{equation}
}
In particular, for $i=n$ due to the fact that $f_{n+2,n}=f_{n+3,n}=0$ we have
\begin{equation}
\begin{split}
	0=&
	\BK{\frac{T}{N}\tr{\ww_1(x_1)\left(V_n^\prime(M_n)-\cc{n-1,n}M_{n-1}\right)Q(z)}}\\
	&-\cc{n,n+1}z W_{n+1}(x_1,z)+S(z)W_0(x_1)
\end{split}
	\label{eq:LE2b}
\end{equation}

\subsection{Master Loop Equation}\label{sSec:MLE}

When we sum up equations \eqref{eq:LE1z} and \eqref{eq:LE2a} for $i=2,\dotsc,n$ we find the \emph{master loop equation}
{\small
\begin{equation}
\begin{split}
	\frac{T^2}{N^2}W_{2;1}(x_1,x_2,\dotsc,x_n,x_{n+1};x_1)+&
	(\cc{1,2}x_2-V_1^\prime(x_1)+W_0(x_1))\left(W_2(x_1,x_2,\dotsc,x_n,x_{n+1})-S(x_{n+1})\right)=\\
	=&-W_1(x_1,\dotsc,x_n,x_{n+1})+(V_1^\prime(x_1)-\cc{1,2}x_2)S(x_{n+1})\\
	&+\sum_{i=2}^n (V_i^\prime(x_i)-\cc{i-1,i}x_{i-1}-\cc{i,i+1}x_{i+1})W_{i+1}(x_1,x_{i+1},\dotsc,x_{n+1})
\end{split}
	\label{eq:MLE}
\end{equation}
}
where we have redefined $z\equiv x_{n+1}$. Remember that $W_1(x_1,\cdots,x_n)$ is a polynomial in all its variables and that $W_i(x_1,x_i,\cdots,x_n)$ is a polynomial in all its variables except $x_1$.
In particular, we may choose $x_i=\hat{x}_i(x_1,x_2)$, $i=3,\dots,n$ such that
\begin{equation}
V'_i(x_{i})=\cc{i-1,i}x_{i-1}+\cc{i,i+1}x_{i+1}\quad , \quad \forall i=2,\dots,n
\label{eq:CC.a}
\end{equation}
and in that case \eqref{eq:MLE} reduces to
\begin{equation}
	\begin{split}
	\frac{T^2}{N^2}\hat{W}_{2;1}(x_1,x_2;x_1)+
	(\cc{1,2}x_2-Y(x_1))\hat{U}(x_1,x_2)=&-\hat{W}_1(x_1,x_2)+(V_1^\prime(x_1)-\cc{1,2}x_2)
	\hat{S}(x_1,x_2)\\
	=&\hat{E}(x_1,x_2)
	\end{split}
	\label{eq:rMLE}
\end{equation}
where we have defined
\begin{equation}
	\begin{split}
	Y(x_1) &= V'_1(x_1)-W_0(x_1) \\
	\hat{U}(x_1,x_2)&=W_2(x_1,x_2,\hat{x}_3,\dotsc,\hat{x}_{n+1})-S(\hat{x}_{n+1})\\
	\hat{W}_{2;1}(x_1,x_2;x_1)&=W_{2;1}(x_1,x_2,\hat{x}_3,\dotsc,\hat{x}_{n+1};x_1)\\
	\hat{W}_1(x_1,x_2)&=W_1(x_1,x_2,\hat{x}_3,\dotsc,\hat{x}_{n+1})\\
	\hat{S}(x_1,x_2)&=S(\hat{x}_{n+1})
	\end{split}
	\label{eq:rMMO}
\end{equation}
and $\hat{x}_i$ are defined recursively from the constraints \eqref{eq:CC.a}
\begin{equation}
	\begin{split}
	\cc{2,3}\hat{x}_3&=\cc{2,3}\hat{x}_3(x_1,x_2)=V_2^\prime(x_2)-\cc{1,2}x_1\\
	\cc{3,4}\hat{x}_4&=\cc{3,4}\hat{x}_4(x_1,x_2)=V_3^\prime(\hat{x}_3(x_1,x_2))-\cc{2,3}x_2\\
	\cc{i-1,i}\hat{x}_i&=\cc{i-1,i}\hat{x}_i(x_1,x_2)=
	V_{i-1}^\prime(\hat{x}_{i-1}(x_1,x_2))-\cc{i-2,i-1}\hat{x}_{i-2}(x_1,x_2)\,,\quad \textrm{for }i>4
	\end{split}
	\label{eq:rV}
\end{equation}
Note the resemblance between equations \eqref{eq:CC.a} and \eqref{eq:FP}.

\subsection{Planar limit}

To leading order at large $N$, we drop the $T^2/N^2$ term in the loop equation \eqref{eq:rMLE} and we get:
\begin{equation}
	(\cc{1,2}x_2-Y^{(0)}(x_1))\hat{U}^{(0)}(x_1,x_2)
	=\hat{E}^{(0)}(x_1,x_2)
	\label{eq:rMLELN}
\end{equation}
Notice that $\hat{E}^{(0)}(x_1,x_2)$ is a polynomial in its 2 variables $x_1$ and $x_2$.\\
The algebraic equation
\begin{equation}
\boxed{ \hat{E}^{(0)}(x_1,x_2) = 0    }
\end{equation}
is called the spectral curve.
In some sense it is the large $N$ limit of the loop equation when we choose $\cc{1,2}x_2=Y^{(0)}(x_1)$.

The equation \eqref{eq:rMLE} is of the same form as the one solved in \cite{ChEyOr} for the 2-matrix model, or the one solved in \cite{EOFg} for the 1-matrix model with external field. It can thus be solved using the same methods. Note that (as we said above) we consider fixed filling fractions in the formal model, which means that the
\begin{equation}
	\epsilon_i =-\frac{1}{2i\pi}\oint_{\A_i} W_0^{(0)}(x_1(p))\d x_1(p)
	\label{eq:FFF}
\end{equation}
are fixed data of the model. As a consequence of that, all the differentials $W_0^{(h)}(x_1(p))\d x_1(p)$ with $h\geq 1$ may have poles only at the branch points $\alpha_i$, and all their $\A$ cycles integrals are zero due to the fixed filling fractions condition
\begin{equation}
	0=\oint_{\A_i} W_0^{(h)}(x_1(p))\d x_1(p).
	\label{eq:Ac0}
\end{equation}

\section{Algebraic geometry of the spectral curve}\label{Sec:AC}

The solution of the model relies on the understanding of the underlying large $N$ spectral curve, and its algebraic-geometry properties. Let us see first what are the main features of $\hat{E}^{(0)}(x_1,x_2)$ and then we will present a set of tools and concepts that we will need later.

\medskip

First, $\hat{E}^{(0)}(x_1,x_2)=-W^{(0)}_1(x_1,x_2,\hat{x}_3,\dotsc,\hat{x}_{n+1})
+(V_1^\prime(x_1)-\cc{1,2}x_2)S(\hat{x}_{n+1})$ as we have noted before, is a polynomial in all its variables.
$W^{(0)}_1$ is a polynomial of degree $d_i-1$ in the variable $x_i$ and $\s-1$ in $z$ while $(V_1^\prime(x_1)-\cc{1,2}x_2)S(x_{n+1})$ is clearly a polynomial of degree $d_1$ in $x_1$, $1$ in $x_2$ and $\s$ in $z=x_{n+1}$.

The relations \eqref{eq:rV} express $\hat{x}_i$ ($i\ge 3$) as a polynomial of $x_1$ and $x_2$.
For example, $\hat{x_3}(x_1,x_2)$ is a polynomial of degree $1$ in $x_1$ and $d_2$ in $x_2$.
In general, for $i>3$, $\hat{x}_i(x_1,x_2)$ a polynomial of degree $\prod_{j=3}^{i-1}d_j$ in $x_1$ and $\prod_{j=2}^{i-1}d_j$ in $x_2$.

With this information we see that for $n>1$, $\hat{W}^{(0)}_1(x_1,x_2)$ is a polynomial of degree
$d_1+d_3\dots d_n \s -2$ in $x_1$ and a polynomial of degree $d_2\dotsm d_{n} \s-1$ in $x_2$, while $(V_1^\prime(x_1)-\cc{1,2}x_2)S(\hat{x}_{n+1})$ is a polynomial of degree $d_1+d_3\dots d_n\, \s$ in $x_1$ and $1+d_2 d_3\dots d_n\, \s$ in $x_2$, i.e.
\begin{equation}
\deg_{x_1} \hat{E}^{(0)} = d_1+d_3\dots d_n\, \s = d_1 + D_1
\quad , \quad
\deg_{x_2} \hat{E}^{(0)} = 1+d_2 d_3\dots d_n\, \s = 1+D_2
	\label{eq:deg}
\end{equation}
One can check from algebraic geometry usual methods (Newton's polytope for instance), that an algebraic curve with those degrees, has  a genus $\genus$:
\begin{equation}
\genus< d_1 d_2 \dots d_n\, \s\,\,
	\label{eq:genus}
\end{equation}

\medskip
So far, most of the coefficients of $\hat{E}^{(0)}$ are not known, because they come from the unknown polynomial $W_1$. However, the number of unknown coefficients of $W_1$, is $d_1 d_2 \dots d_n\, \s\,\,-1$, and it matches precisely the generic genus of the spectral curve $\genus$ (all the terms of $W_1$ lie in the interior of Newton's polytope), and therefore the polynomial $W_1$ (and thus $\hat{E}^{(0)}$) is entirely determined by the filling fraction conditions (we have $\sum_i \epsilon_i=T$):
\begin{equation}
	\forall i=1,\dots,d_1 d_2 \dots d_n\, \s, \quad \epsilon_i = \frac{1}{2i\pi} \oint_{\A_i} Y(x)dx
	\qquad , \quad \hat{E}^{(0)}(x,Y(x))=0
	\label{eq:FFcond}
\end{equation}
Those $d_1 d_2 \dots d_n\, \s$ equations determine $W_1^{(0)}$ and thus $\hat{E}^{(0)}$.

\subsection{Analytical structure, sheets and poles}

The algebraic curve $\hat{E}^{(0)}(x_1,x_2)=0$ has the following structure.
For each value of $x_1$ there are $D_2+1$ different values of $x_2$, and  for every value of $x_2$ we find $D_1+d_1$ values of $x_1$. This observation is what defines, respectively, the $x_1$ sheet structure and the $x_2$ sheet structure.

The algebraic curve $\hat{E}^{(0)}(x_1,x_2)=0$ can be parametrized as follows:
there exists a compact Riemann surface ${\cal L}$ and two meromorphic functions $x_1$ and $x_2$ on ${\cal L}$, such that
\begin{equation}
\hat{E}^{(0)}(x_1,x_2)=0
\quad \Leftrightarrow
\exists p\in {\cal L}\,\,\, | \,\,\, x_1=x_1(p)\, \text{ and }\, x_2=x_2(p)
\end{equation}
Notice that the functions $x_i(p)=\hat{x}_i(x_1(p),x_2(p))$ are also meromorphic functions on ${\cal L}$, which satisfy:
\begin{equation}
\forall \, p\in {\cal L}\, , \quad V'_i(x_i(p)) = \cc{i-1,i}x_{i-1}(p)+\cc{i,i+1}x_{i+1}(p) 
\end{equation}
There are $\s+1$ different points on the curve where $x_1$ (and all the other $x_i(p)$, $i\le n$) becomes infinite. Around one of these points, that we will call $p=\infty$, a good local coordinate is $z_\infty(p)=\frac{1}{x_1(p)}$. The $\infty$ point in the curve is quite important as it marks the so called physical sheet. The other "infinity" points correspond to the situation where $x_{n+1}(p)=\lambda_i$, and will be called $p=\hat{\lambda}_i$. A good local coordinate around these points is $z_{\hat{\lambda}_i}(p)=\frac{1}{x_n(p)}$ (a different good local coordinate could be $z_{\hat{\lambda}_i}(p)=x_{n+1}(p)-\lambda_i$ which behaves as $\simif_{p\to \lambda_i} \frac{1}{\cc{n,n+1}}\frac{Tl_i}{N}\frac{1}{x_n(p)}$).
Explicitly, the negative divisor of $x_k(p)$ is
\begin{equation}
\begin{split}
	[x_k(p)]_-&= -r_k \infty- s_k\sum_i \hat{\lambda}_i \\
\text{where     }& r_1=1,\,\,\, r_k=d_1 d_2\dots d_{k-1} \\
& s_{n+1}=0,\,\,\, s_n=1,\,\,\, s_k=d_{k+1} d_{k+2}\dots d_{n} \\
\end{split}
	\label{eq:Div}
\end{equation}
Locally, near $\infty$ we have:
\begin{equation}
x_2(p) \eqif_{p\to\infty} \frac{V'_1(x_1(p))}{\cc{1,2}} - \frac{T}{x_1(p)} + \O{x_1(p)^{-2}}
\end{equation}
\begin{equation}
x_k(p) \simif_{p\to\infty} x_1(p)^{r_k}
\end{equation}
and near $\hat\lambda_i$:
\begin{equation}
\begin{split}
x_n(p) &\eqif_{p\to\lambda_i} \frac{1}{\cc{n,n+1}}\frac{Tl_i}{N}\,\,\frac{1}{x_{n+1}(p)-\lambda_i} +\O{1}\\
x_{n+1}(p) &\eqif_{p\to\lambda_i} \lambda_i + \frac{1}{\cc{n,n+1}}\frac{Tl_i}{N}\, \frac{1}{x_n(p)} + \O{x_n(p)^{-2}}
\end{split}
\end{equation}
\begin{equation}
x_k(p) \simif_{p\to\hat\lambda_i} (x_{n+1}(p)-\lambda_i)^{-s_k}
\end{equation}

\subsection{Branchpoints and conjugated points}

From Riemann-Hurwitz, there are $\s+2\genus+s_1$ points $\alpha_i$ on ${\cal L}$, such that $\partial_{x_2} \hat{E}^{(0)}(x_1,x_2)=0$ and $\partial_{x_1} \hat{E}^{(0)}(x_1,x_2)\neq 0$. They are called the $x_1$ branch points. They are the zeros of the differential $dx_1(p)$.

For the moment, we assume that the branch points are simple, i.e. that 
at those points $\d x_1(p)$ vanishes linearly when $p\to\alpha$.
The spectral curve is said to be regular. 
A spectral curve with non simple branch-points is called singular or critical.
We study critical points below in section \ref{Sec:DSL}.

\medskip

Assuming that the spectral curve is regular means that near any branch-point $\alpha$, $Y=\cc{1,2}x_2$ behaves locally like a square root $Y(x_1)\sim Y(x_1(\alpha))+ C\,\sqrt{x_1-x_1(\alpha)}$, and therefore, for any $p$ in the vicinity of $\alpha$, there exists a unique point $\bar{p}\neq p$ in the same vicinity of $\alpha$, such that
\begin{equation}
x_1(\bar{p})=x_1(p).
\end{equation}
We say that $\bar{p}$ is the conjugate point of $p$. The conjugate point, is defined locally near every branch-point, and in general it is not defined globally (see \cite{EOFg}).

\subsection{Non-trivial cycles}

If ${\cal L}$ is of genus $\genus$, there exists a symplectic basis of non-trivial cycles ${\cal A}_i,{\cal B}_j$, $i,j=1,\dots,\genus$, such that:
\begin{equation}
{\cal A}_i\cap{\cal B}_j=\delta_{i,j}
\quad , \quad
{\cal A}_i\cap{\cal A}_j=0
\quad , \quad
{\cal B}_i\cap{\cal B}_j=0
\end{equation}
Such a basis is not unique, and we have to choose one of them. Different choices give different solutions of the loop equations. The choice is related to the choice of filling fractions.

\medskip

Changes of symplectic basis are called modular transformations, and, following \cite{EOFg, EOMham} we study modular transformations of the $F_g$'s and $W_0^{(h)}$'s in section \ref{Sec:MT}.

\bigskip

Once we have chosen a basis of non-trivial cycles, the domain ${\cal L}\backslash (\cup_{i} \A_i \cup_i \B_i)$ is simply connected and is called the fundamental domain.

\subsection{Bergman kernel}

We use the notations of \cite{EOFg}, and we refer the reader to \cite{EOFg} for a more detailed description. 

On every compact Riemann surface ${\cal L}$, with a given symplectic basis of non trivial cycles , is defined uniquely a 2nd kind differential called the Bergman kernel \cite{BergSchif} $B(p_1,p_2)$ (which we regard as a 2nd kind differential in the variable $p_1\in {\cal L}$), that satisfies
\begin{itemize}
	\item[{\it i)}]{$B(p_1,p_2)$ has a double pole, with no residue, when $p_1\to p_2$, and normalized such that
\begin{equation}
B(p_1,p_2) \simif_{p_1\to p_2} \frac{dx(p_1)dx(p_2)}{(x(p_1)-x(p_2))^2} + {\rm finite}
\end{equation}
where $x(p)$ can be any local parameter in the vicinity of $p_2$.
}
\item[{\it ii)}]{$$\oint_{{\cal A}_i} B(p_1,p_2)=0.$$
}
\end{itemize}

It is easy to see that the Bergman kernel is unique, because the difference of 2 Bergman kernels would have no pole and vanishing ${\cal A}-$cycle integrals, i.e. it would vanish.

More explicitly we have:
\begin{equation}
B(p_1,p_2)  = d_{p_1}\, d_{p_2}\,\log{(\theta(u(p_1)-u(p_2)-\kappa))}
\end{equation}
where $\theta$ is the theta-function, $u(p)$ is the Abel map, and $\kappa$ is some odd characteristics.

\medskip

For example, in the case the spectral curve has genus zero (the so-called 1-cut case), ${\cal L}$ is the Riemann sphere, i.e. the complex plane with a point at $\infty$, and $B(p_1,p_2)$ is the meromorphic  bilinear form
$B(p_1,p_2)=\frac{dp_1\, dp_2}{(p_1-p_2)^2}$.
Another example is the case where ${\cal L}$ is a torus of modulus $\tau$: ${\cal L}={\mathbb C}/({\mathbb Z}+\tau{\mathbb Z})$, for which the Bergman kernel is the Weierstrass function:
$B(p_1,p_2) = (\wp(p_1-p_2;\tau)+C) dp_1 dp_2$.

\subsection{Third kind differential}

For any $p\in {\cal L}$ and two points $q$ and $o$ in the fundamental domain, we define:
\begin{equation}
dS_{q,o}(p) = \int_o^q B(p,q')
\end{equation}
where the integration contour lies in the fundamental domain (i.e. it does not intersect any $\A$ or $\B$ cycles).
$dS_{q,o}(p)$ is a meromorphic differential form in the variable $p$, whereas it is a scalar function of $q$ and $o$.
It has a simple pole at $p=q$ with residue $+1$ and a simple pole at $p=o$ with residue $-1$:
\begin{equation}
\mathop{{\rm Res}}_{p\to q}\,dS_{q,o}(p) = +1
\qquad , \qquad
\mathop{{\rm Res}}_{p\to o}\,dS_{q,o}(p) = -1
\end{equation}
i.e. it behaves locally like $\frac{dx(p)}{x(p)-x(q)}$ when $p\to q$, in any local parameter $x(p)$.
Moreover it has vanishing $\A$ cycle integrals:
\begin{equation}
\oint_{\A_i}\,dS_{q,o} = 0
\end{equation}
Since it has only one simple pole in the variable $q$, this 3rd kind differential is very useful for writing Cauchy residue formula. For any meromorphic differential form $\omega(p)$ we have:
\begin{equation}\label{eqresCauchydS1}
\omega(p) = -\mathop{{\rm Res}}_{q\to p}\,dS_{q,o}(p) \,\omega(q)
\end{equation}
and, using Riemann bilinear identity \cite{Farkas, Fay}, if $\forall i, \oint_{\A_i} \omega=0$, and $\omega$ has poles $\alpha_i$'s, we may move the integration contour and get:
\begin{equation}\label{eqresCauchydS}
\omega(p) = \sum_i \mathop{{\rm Res}}_{q\to \alpha_i}\,dS_{q,o}(p) \,\omega(q)
\end{equation}
This identity was the main ingredient in solving loop equations for the 1-matrix model in \cite{eyn1loop}.

\section{Solution of the Loop Equation}\label{Sec:SMLE}

In this section we solve the loop equation to all orders in the topological $T^2/N^2$ expansion.

We first need a technical lemma which consists in proving that the solution is unique, and then we use this uniqueness to try a guess similar to that introduced in \cite{ChEyOr} which makes the loop equations easier to solve.

We find the one point resolvent and the $k$ point resolvent for the first matrix of the chain, and in fact we find that they coincide with the correlators defined in \cite{EOFg} for the spectral curve ${\hat E}^{(0)}$.

\subsection{Unicity of the solution}\label{Sec:UotS}

Equation \eqref{eq:rMLE} fixes the large $N/T$ expansion of $W_{0}(x_1(p))$\footnote{As we show later all the $k$-functions of the type
\begin{equation}
	W_{0;1^{k-1}}^{(h)}(x_1(p),x_1(q^{(l)}),\dotsc,x_1(p^{(l)}))=
	\left(\prod_{l=1}^{k-1}\frac{\partial}{\partial V_1^\prime(x_1(q^{(l)}))}\right) W_{0}^{(h)}(x_1(p))\nonumber
\end{equation}
can be determined from the equation \eqref{eq:rMLE} exactly in the same way as in \cite{EOFg,ChEyOr}$\dots\,\,\,$.}.
Take equation \eqref{eq:rMLE} and substitute the $\frac{T^2}{N^2}$ expansion of $\hat{W}_{2;1}(x_1,x_1^\prime)$, $W_0(x_1)$, $\hat{U}(x_1,x_2)$ and $\hat{E}(x_1,x_2)$. Then to order $\frac{T^{2h}}{N^{2h}}$ we obtain
\begin{equation}
\begin{split}
	(\cc{1,2}x_2-Y(x_1))\hat{U}^{(h)}(x_1,x_2)+W_0^{(h)}(x_1)\hat{U}^{(0)}(x_1,x_2)=&\\
	=\hat{E}^{(h)}(x_1,x_2)-\sum_{m=1}^{h-1}W_0^{(h-m)}(x_1)
    &\hat{U}^{(m)}(x_1,x_2)-\hat{W}_{2;1}^{(h-1)}(x_1,x_2;x_1).
\end{split}
    \label{eq:ghMLE.a}
\end{equation}
Suppose you know $\hat{U}^{(h^\prime)}(x_1,x_2)$, $W_0^{(h^\prime)}(x_1)$ and $\hat{E}^{(h^\prime)}(x_1,x_2)$ for $h^\prime<h$. We prove that we can find those three functions for $h^\prime=h$.
Consider $x_1=x_1(q)$ and $x_2=x_2(p)$ (and so $Y(x_1(q))=\cc{1,2}x_2(q)$) with $p$ and $q$ living on the algebraic curve
\begin{equation}
\begin{split}
	\cc{1,2}(x_2(p)-x_2(q))\,\hat{U}^{(h)}(x_1(q),x_2(p))+W_0^{(h)}(x_1(q))\hat{U}^{(0)}(x_1(q),x_2(p))=&\\
	=\hat{E}^{(h)}(x_1(q),x_2(p))
	-\sum_{m=1}^{h-1}W_0^{(h-m)}(x_1(q))\hat{U}^{(m)}(x_1(q),x_2(p))-
    &\hat{W}_{2;1}^{(h-1)}(x_1(q),x_2(p);x_1(q)).
\end{split}
	\label{eq:ghMLE.b}
\end{equation}
Begin with $h=0$. Consider the solutions for the equations $\hat{E}(x_1(q),x_2(p))=0$. For every $x_1(q)$ there are $D_2+1$ different solutions $Y(x_1(q^{(i)}))$ (sitting at points that we call $q^{(0)},q^{(1)},\dotsc,q^{{(D_2)}}$ on the curve, with the convention that $q^{(0)}=q$). Then we can write
\begin{equation}
\begin{split}
	\hat{E}^{(0)}(x_1(q),x_2(p))&=K\prod_{i=0}^{D_2}\left(\cc{1,2}x_2(p)-Y(x_1(q^{(i)}))\right)\\
	\hat{U}^{(0)}(x_1(q),x_2(p))&=\frac{\hat{E}^{(0)}(x_1(q),x_2(p))}{(\cc{1,2}x_2(p)-Y(x_1(q)))}
	=K\prod_{i=1}^{D_2}\left(\cc{1,2}x_2(p)-Y(x_1(q^{i}))\right)
\end{split}
	\label{eq:US.a}
\end{equation}
where the constant $K$ is derived in the next section.
Recall that $x_1(q^{(i)})=x_1(q^{(j)})$ but in general\footnote{The function $Y(x_1(q^{(i)}))$ is multi valued in the $x_1$ plane. On the other side on the algebraic curve it is not multi valued. The index $i$ indicates precisely different $x_1$-sheets, and thus different values of $Y(x_1)$.} $Y(x_1(q^{(i)}))\not=Y(x_1(q^{(j)}))$ for $i\not=j$.

\medskip

Consider now $h>0$.\\
Write now the equation for an arbitrary $h$ and take $p\to q^0=q$
\begin{equation}
	\begin{split}
	W_0^{(h)}(x_1(q))&\hat{U}^{(0)}(x_1(q),x_2(q))=\hat{E}^{(h)}(x_1(q),x_2(q))\\
	& -\sum_{m=1}^{h-1}W_0^{(h-m)}(x_1(q))\hat{U}^{(m)}(x_1(q),x_2(q))-\hat{W}_{2;1}^{(h-1)}(x_1(q),x_2(q);x_1(q)).
	\end{split}
	\label{eq:ghMLE.c}
\end{equation}
This equation shows (by recursion) that $W_0^{(h)}(x_1(q))$ is a meromorphic function on the spectral curve, and because of our hypothesis, it has poles only at branch-points, and it has vanishing $\A$ cycle integrals.
Let us write Cauchy residue formula \eqref{eqresCauchydS1}:
\begin{equation}
W_0^{(h)}(x_1(q)) dx_1(q)  = - \Res_{q'\to q}\, dS_{q',o}(q)\,\, W_0^{(h)}(x_1(q')) dx_1(q')  
\end{equation} 
Using Riemann bilinear identity, and the fact that both $dS$ and $W_0^{(h)}dx$ have vanishing $\A$ cycle integrals, we can move the integration contour and get \eqref{eqresCauchydS}:
\begin{equation}
W_0^{(h)}(x_1(q)) dx_1(q)  = \sum_\alpha \Res_{q'\to \alpha}\, dS_{q',o}(q)\,\, W_0^{(h)}(x_1(q')) dx_1(q')  
\end{equation} 
Now we replace $W_0^{(h)}(x_1(q'))$ in the RHS with the loop equation \ref{eq:ghMLE.c}, and using that $\hat{E}^{(h)}(x_1,x_2)$ is a polynomial and has no poles at finite $x_1$, that $\hat{U}^{(0)}(x_1(q),x_2(q))$ vanishes at most as a square root at the branch points and that $\d{x_1(p)}$ vanishes linearly at the branchpoints, we find:
\begin{equation}
	\begin{split}
	&W_0^{(h)}(x_1(q))\d x_1(q)=\sum_{\alpha}\Res_{q^\prime\to\alpha}W_0^{(h)}(x_1(q^\prime))\d x_1(q^\prime)
	\d S_{q^\prime,o}(q)=\\
	&\hspace{50pt}=-\sum_{\alpha}\Res_{q^\prime\to\alpha}
	\frac{\d x_1(q^\prime)\d{}	S_{q^\prime,o}(q)}{\hat{U}^{(0)}(x_1(q'),x_2(q'))}
	\Bigg(	\sum_{m=1}^{h-1}W_0^{(h-m)}(x_1(q'))\hat{U}^{(m)}(x_1(q'),x_2(q'))\\
	&\hspace{250pt}+\hat{W}_{2;1}^{(h-1)}(x_1(q'),x_2(q');x_1(q'))\Bigg).
	\end{split}
	\label{eq:ghMLE.d}
\end{equation}
where everything on the RHS is known from the recursion hypothesis, and thus determine uniquely $W_0^{(h)}(x_1(q))$.

Then, consider again equation \eqref{eq:ghMLE.c} and find $\hat{E}^{(h)}(x_1(q),x_2(q))$ (equal to $\hat{E}^{(h)}(x_1(q^{(i)}),x_2(q))$ by the definition of $q^{(i)}$)
\begin{equation}
\begin{split}
	 \hat{E}^{(h)}(x_1(q^{(i)}),x_2(q))=\hat{E}^{(h)}&(x_1(q),x_2(q))=W_0^{(h)}(x_1(q))\hat{U}^{(0)}(x_1(q),x_2(q))\\
	 &-\sum_{m=1}^{h-1}W_0^{(h-m)}(x_1(q))\hat{U}^{(m)}(x_1(q),x_2(q))-\hat{W}_{2;1}^{(h-1)}(x_1(q),x_2(q);x_1(q))
\end{split}
	\label{eq:ghMLE.f}
\end{equation}
and reconstruct $\hat{E}^{(h)}(x_1(q),x_2(p))$ using the Lagrange interpolation formula
\begin{equation}
	\hat{E}^{(h)}(x_1(q),x_2(p))=\sum_{i} \frac{\hat{E}^{(h)}(x_1(q),x_2(q^{(i)}))\prod_{j\not= i}(x_2(q^j)-x_2(p))}
	{\prod_{j\not= i}(x_2(q^j)-x_2(q^{(i)}))}.
	\label{eq:IF}
\end{equation}
Finally equation \eqref{eq:ghMLE.b} gives $\hat{U}^{(h)}(x_1(q),x_2(p))$.

Therefore we have proved our recursion hypothesis to order $h$.

All this procedure allows us to solve recursively the master loop equation, thus indicating that the solution is unique once $E^{(0)}(x_1,x_2)$ (or equivalently $Y(x_1(p))$ and $x_1(p)$) is given. 
We could iterate this procedure indefinitely. We now show a much better way to solve the master loop equation.

\subsection{Solution of the equation}\label{Sec:SofE}

The solution being unique, we only have to find one solution. 
The equation
\begin{equation}
	\begin{split}
	\frac{T^2}{N^2}\hat{W}_{2;1}(x_1,x_2;x_1)+&
	(\cc{1,2}x_2-V_1^\prime(x_1)+W_0(x_1))\hat{U}(x_1,x_2)=\\
	=&-\hat{W}_1(x_1,x_2)+(V_1^\prime(x_1)-\cc{1,2}x_2)\hat{S}(x_1,x_2)=\hat{E}(x_1,x_2)
	\end{split}
	\label{eq:rMLE.b}
\end{equation}
is indeed solved by the expressions
\begin{equation}
\begin{split}
	\hat{E}(x_1(p),x_2)&=-K "\BK{\prod_{i=0}^{D_2}
	\left(\cc{1,2}x_2-V_1^\prime(x_1(p))+\frac{T}{N}\tr{\frac{1}{x_1(p^{(i)})-M}}\right)}"\\
	&\hat{U}(x_1(p),x_2)=
	-K "\BK{\prod_{i=1}^{D_2}\left(\cc{1,2}x_2-V_1^\prime(x_1(p))+\frac{T}{N}\tr{\frac{1}{x_1(p^{(i)})-M}}\right)}".
\end{split}
	\label{eq:SrMLE}
\end{equation}
and can be proved following \cite{ChEyOr}.
The product runs over the $D_2+1$ sheets of the algebraic curve viewed from the $x_1$ variable point of view. The $0$th sheet is by definition the sheet in which the point $p$ is sitting (that is, $p=p^{(0)}$). The notation $"\BK{\dotsm}"$ means that if we expand the product in cumulants, the connected two point correlators must be replaced by $\overline{W}_{0;1}(x_1(p);x_1(p^\prime))= W_{0;1}(x_1(p);x_1(p^\prime))+\frac{1}{(x_1(p)-x_1(p^\prime))^2}$. 

\medskip

These expressions are not of practical immediate use, but if we expand them in powers of $x_2$ they reveal the equation that lead us to the explicit solution. All the information is contained in the highest powers.

\begin{itemize}
	\item {$\left(\cc{1,2}x_2\right)^{D_2+1}$:}
\begin{equation}
	\left(\frac{1}{\cc{1,2}}\right)^{d_2\dotsm d_n\s}
	\left(\frac{g_{d_2+1}^{(2)}}{\cc{2,3}}\right)^{d_3\dotsm d_n\s}
	\left(\frac{g_{d_3+1}^{(3)}}{\cc{3,4}}\right)^{d_4\dotsm d_n\s}
	\dotsm 
	\left(\frac{g_{d_n+1}^{(n)}}{\cc{n,n+1}}\right)^{\s}=K
	\label{eq:D2+2}
\end{equation}

	\item {$\left(\cc{1,2}x_2\right)^{D_2}$:}
\begin{equation}
\begin{split}
	K	\left[V_1^\prime(x_1)-d_3\dotsm d_n\s\frac{\cc{1,2}g_{d_2}^{(2)}}{g_{d_2+1}^{(2)}}\right]&=
	K \sum_{i=0}^{D_2}\left(V_1^\prime(x_1(p))-\BK{\frac{T}{N}\tr{\frac{1}{x_1(p^{(i)})-M}}}\right)\\
	V_1^\prime(x_1)-d_3\dotsm d_n\s\frac{\cc{1,2}g_{d_2}^{(2)}}{g_{d_2+1}^{(2)}}&=\sum_{i=0}^{D_2} Y(x_1(p^{(i)}))
\end{split}
	\label{eq:D2+1}
\end{equation}
where we have defined as usual $V^\prime_1(x_1)-W_0(x_1(p^{(i)}))= Y(x_1(p^{(i)}))$
	\item {$\left(\cc{1,2}x_2\right)^{D_2-1}$:
\begin{equation}
\begin{split}
	&P(x_1)-d_3\dotsm d_n\s\frac{\cc{1,2}g_{d_2}^{(2)}}{g_{d_2+1}^{(2)}}\left(V_1^\prime(x_1)
	-(d_3\dotsm d_n\s -1)\frac{1}{2}\frac{\cc{1,2}g_{d_2}^{(2)}}{g_{d_2+1}^{(2)}}
	-\frac{\cc{1,2}g_{d_2-1}^{(2)}}{g_{d_2}^{(2)}}\right)=\\
	&\hspace*{3cm}=\frac{1}{2}\sum_{i\not=j}^{D_2}
	 \left(Y(x_1(p^{(i)}))Y(x_1(p^{(j)}))+\frac{T^2}{N^2}\overline{W}_{0;1}(x_1(p^{(i)}),x_1(p^{(i)}))\right)
\end{split}
	\label{eq:D2}
\end{equation}
where $P(x_1)=\Pol_{x_1}V_1^\prime(x_1)W_0(x_1)$ was already defined in \eqref{eq:MMO} and
\begin{equation}
	\overline{W}_{0;1}(x_1(p),x_1(q))={W}_{0;1}(x_1(p),x_1(q))+\frac{1}{(x_1(p)-x_1(q))^2}.
	\label{eq:Wbar}
\end{equation}
refers to the substitution mentioned above for two point correlators.}
\end{itemize}
The equation \eqref{eq:D2+2} allows us to determine the constant $K$. The equation \eqref{eq:D2+1} allows us to modify the last equation.
When doing the $T^2/N^2$ expansion, equation \eqref{eq:D2+1} implies
\begin{equation}
	\sum_{i=0}^{D_2}W_0^{(h)}(x_1(p^{(i)}))=0 \quad\text{for } h>0
	\label{eq:D2+1.b}
\end{equation}
Apply $\frac{\partial}{\partial V_1(x(q))}$ to equation \eqref{eq:D2+1} we find also
\begin{equation}
	\sum_{i=0}^{D_2} \overline{W}^{(h)}_{0;1}(x_1(p^{(i)}),x_1(q))=\delta_{h,0}\frac{1}{(x_1(q)-x_1(p))^2}
	\label{eq:D2+1.c}
\end{equation}
Using all these equations we find that equation \eqref{eq:D2} can be transformed into
\begin{equation}
\begin{split}
	 \sum_{i=0}^{D_2}&\left[{Y(x_1(p^{(i)}))}^2+\frac{T^2}{N^2}W_{0;1}(x_1(p^{(i)}),x_1(p^{(i)}))\right]=\\
	&\qquad =\left(V_1^\prime(x_1)\right)^2-P(x_1)-
	d_3\dots d_n\s\left(d_3\dots	 d_n\s-2\right)\left(\frac{\cc{1,2}g_{d_2}^{(2)}}{g_{d_2+1}^{(2)}}\right)^2
	-2d_3\dotsm d_n\s\frac{{\cc{1,2}}^2g_{d_2-1}^{(2)}}{g_{d_2+1}^{(2)}}
\end{split}
	\label{eq:D2.b}
\end{equation}
Expanding the equation in $\frac{T}{N}=\hbar$ as in \cite{ChEyOr} we get for $h\geq 1$ the equation (with $y(p)=Y(x_1(p))$)
\begin{equation}
\begin{split}
	2\sum_{i=1}^{D_2}y(p^{(i)})W_{0}^{(h)}(x_1(p^{(i)}))=&\\
	=\sum_{i=1}^{D_2}\sum_{m=1}^{h-1}W_{0}^{(m)}(x_1(p^{(i)}))W_{0}^{(h-m)}(x_1(p^{(i)}))&+
	\sum_{i=1}^{D_2}W_{0;1}^{(h-1)}(x_1(p^{(i)}),x_1(p^{(i)}))+ 2P^{(h)}(x_1)
\end{split}
	\label{eq:hRR}
\end{equation}
The rest follows exactly the same lines as in \cite{ChEyOr}.
We will however recall the main steps.
Let us define the following meromorphic differentials from the correlation functions
\begin{equation}
	\w_{k}^{(h)}(p_1,\dotsc,p_k)=
	\left(\prod_{j=1}^{k}\d{x_1(p_j)}\right)
	\left(\prod_{i=2}^{k}\frac{\pd{}}{\pd{}V_1(x_1(p_i))}\right) W_{0}^{(h)}(x_1(p_1))
	\label{eq:CF2DF}
\end{equation}
and rewrite equation \eqref{eq:hRR} as
\begin{equation}
\begin{split}
	2\sum_{i=1}^{D_2}y(p^{(i)}) \w_{1}^{(h)}(p^{(i)}) \d{x_1(p^{(i)})}=&\\
	=\sum_{i=1}^{D_2}\sum_{m=1}^{h-1}\w_{1}^{(m)}(p^{(i)})\w_{1}^{(h-m)}(p^{(i)})&+
	\sum_{i=1}^{D_2}\w_{2}^{(h-1)}(p^{(i)},p^{(i)})+ 2P^{(h)}(x_1(p))\d{x_1(p)}^2
\end{split}
	\label{eq:DhRR}
\end{equation}
Define also the third kind differential $\d{E_{p,\bar{p}}}(q)=\d{S_{p,o}}(q)-\d{S_{\bar{p},0}}(q)$, where $\bar{p}$ is the conjugated point of $p$. 
Finally apply the operator $\sum_{\alpha}\Res_{p\to\alpha}\frac{1}{2}
\frac{\d{E_{p,\bar{p}}(q)}}{y(x_1(p))-y(x_1(\bar{p}))}$ (where $\alpha$ are the branch points of the curve) to the equation \eqref{eq:CF2DF}. After some algebra we find
\begin{equation}
	\w_1^{(h)}(q)=-\sum_{\alpha}\Res_{p\to\alpha}\frac{1}{2}\frac{\d{E_{p,\bar{p}}(q)}}{(y(p)-y(\bar{p}))
	\d{x_1(p)}}\left(\sum_{m=1}^{h-1}\w_{1}^{(m)}(p)\w_{1}^{(h-m)}(\bar{p})
	+\w_{2}^{(h-1)}(p,\bar{p})\right)	
	\label{eq:hRR.a}
\end{equation}
which is the first of a tower of recursion relations. The rest can be obtained by applying the loop insertion operator to this first one and reads
\begin{equation}
\begin{split}
	\w_{k+1}^{(h)}(q,\{p_K\})&=
	-\sum_{\alpha}\Res_{p\to\alpha}\frac{1}{2}\frac{\d{E_{p,\bar{p}}(q)}}{(y(p)-y(\bar{p}))\d{x_1(p)}}
	\Bigg(\w_{k+2}^{(h-1)}(p,\bar{p},\{p_K\})\\
	&\hspace{100pt}+\sum_{m=0}^{h}\sum_{J\subset K}\w_{j+1}^{(m)}(p,\{p_J\})
	\w_{k+1-j}^{(h-m)}(\bar{p},\{p_{K\backslash J}\})\Bigg)	
\end{split}
	\label{eq:hRR.b}
\end{equation}
where $\{p_K\}$ is a collective notation for $k$ points on the curve, and $K=\{1,\dots,k\}$ is the set of indices. In the expression, $J$ stands for a subset of $j$ elements of $K$, $K\backslash J$ for the complement of $J$ in $K$ and the sum over $J$ and $m$ counts all different subsets and genus, except $(J,m)=(\emptyset,0)$ and $(J,m)=(K,h)$.

\medskip
Therefore we have found that the meromorphic differentials $\w_{k}^{(h)}(q_1,\dots,q_k)$ satisfy exactly the same recursion structure as those of \cite{EOFg}, thus all manipulations done in \cite{EOFg} and other references therein that depend only on this resursion structure need not be repeated here and can be taken as a fact.
For instance it was shown in \cite{EOFg} that the recursion for the $F_g$'s and $W_n^{(g)}$'s can be represented digrammaticaly, and so the same happens here.

\section{Moduli of the chain of matrices and topological expansion of the free energy}\label{Moduli}

In order to find the free energy it is important to understand which are the moduli of the chain of matrices, and  how they change when we change the curve (always within the matrix chain moduli space).

\subsection{Moduli of the chain of matrices}\label{sSec:MCoM}

The chain of matrices is completely characterized by the potentials $V_1(x),\dotsc,V_n(x)$, the interaction parameters $\cc{i,i+1}$\footnote{Note that $\cc{n,n+1}$ can be absorbed in a redefinition of $\Lambda$. We will see later how this appears in the moduli variations.}, the temperature parameter $T$, the eigenvalues and multiplicities of $\Lambda$ and the filling fractions $\epsilon_i$.

It is clear that we can express all these parameters in terms of the meromorphic functions on the curve $x_1(p),\dotsc,x_n(p)$ as follows (with notations borrowed from \cite{Marco2, EOFg, MarcoF}):
\begin{equation}
\begin{split}
	\epsilon_i=&\frac{1}{2\pi i}\oint_{{\cal A}_i}\cc{1,2}x_2 \d{x_1} =\frac{1}{2\pi i}\oint_{{\cal  	A}_i}\cc{k,k+1}x_{k+1}\d{x_k} = - \frac{1}{2\pi i}\oint_{{\cal A}_i}\cc{k,k+1}x_k\d{x_{k+1}} \\
	t_\infty=&T=\Res_{\infty} \cc{1,2}x_2 \d{x_1}=\Res_{\infty} \cc{k,k+1}x_{k+1}\d{x_k}
	=- \Res_{\infty} \cc{k,k+1}x_k\d{x_{k+1}}\\
	t_{\hat{\lambda}_i}=&-\frac{l_i}{N}T
	=-\Res_{\hat{\lambda}_i} \cc{n,n+1}x_n\d{x_{n+1}}
	=-\Res_{\hat{\lambda}_i} \cc{k,k+1}x_{k}\d{x_{k+1}}\\
	&=\Res_{\hat{\lambda}_i} \cc{k,k+1}x_{k+1}\d{x_{k}}
	=\Res_{\hat{\lambda}_i} \cc{1,2}x_{2}\d{x_{1}}  \\
	g_j^{(1)}=&\cc{1,2}\Res_{\infty} x_{1}^{-j}x_{2}\d{x_1} \\
	j>2:\,\,\,\, g_j^{(k)}=&\cc{k,k+1}\Res_{\infty} x_{k}^{-j}x_{k+1}\d{x_k} = \cc{k-1,k}\Res_{\hat\lambda_i} 
	x_{k}^{-j}x_{k-1}\d{x_k} \quad \forall i \\
	t_{\hat{\lambda}_i}\lambda_i=&
	\,-t_{\hat{\lambda}_i}x_{n+1}(\hat{\lambda}_i)=-\cc{n,n+1}\Res_{\hat{\lambda}_i} 
	x_{n+1}(p) x_n(p) \d{x_{n+1}(p)}
\end{split}
	\label{eq:MCoM.1}
\end{equation}
Something deserves attention here: note that $g_j^{(k)}$ can be expressed in $\s+1$ different ways by changing which pole $\hat{\lambda}_i$ or $\infty$ we consider. As we will see later, in order to stay within the matrix chain moduli space, any variation of the curve around one of these points should bring associated other variations around the other points so that the new $g$'s can still be obtained from any of them. Also note that we have not specified how to obtain $\cc{i,i+1}$. They appear in the other equations as to indicate that they are free to choose. Indeed these parameters can always be absorbed into the other parameters of the model (as the equations above indicate). It can also be viewed as a rescaling of the meromorphic functions $x_i(p)$.

Now, we study how the spectral curve changes when we change these parameters (or vice versa).

\noindent Let us define the variations $\Omega$ of the curve by their effect on the differential $\cc{1,2}x_2(p)\d{x_1(p)}$.
Variations of functions or forms, are defined with respect to some fixed variable. There is a Poisson-like structure (thermodynamic identity) indicating how to relate variations with respect to different fixed parameters. The meromorphic form $\Omega$ is defined as:
\begin{equation}
	\left.\delta_\Omega\left(\cc{1,2}x_2(p)\d{x_1(p)}\right)\right|_{x_1(p)}=
	\left.\delta_\Omega(\cc{1,2}x_2(p))\right|_{p}\d{x_1(p)}-\left.\delta_\Omega(x_1(p))\right|_{p}\cc{1,2}\d{x_2(p)}
	=-\Omega(p)
	\label{eq:dO}
\end{equation}
In general we want $\Omega$ to be written in the form
\begin{equation}
	\Omega(p)=\int_{\pd{}\Omega}B(p,q)\Lambda(q)
	\label{eq:Om}
\end{equation}
where $\pd{}\Omega$ is a path which does not intersect circles around branch points.

\subsection{Variation of filling fractions}\label{sSec:Vff}
For variations of the filling fractions we choose 
\begin{equation}
	\Omega(p)=-2i\pi \d{u}_j(p)=-\oint_{{\cal B}_j}B(p,q)
	\label{eq:Vff}
\end{equation}
so that $\pd{}\Omega={\cal B}_j$ and $\Lambda(q)=-1$. From \eqref{eq:MCoM.1} we have
\begin{equation}
	\delta_{\Omega}\epsilon_l=\delta_{jl}\,,\quad\delta_{\Omega}t_\alpha=0\,,\quad\delta_{\Omega}g_j^{(m)}=0\,,\quad
	\delta_{\Omega}\lambda_i=0
	\label{eq:Vff.2}
\end{equation}
so that indeed, $\delta_{-2i\pi\d{u_j}}=\frac{\pd{}}{\pd{}\epsilon_j}$.
Using Theorem 5.1 in \cite{EOFg} we can write
\begin{equation}
	\frac{\pd{}}{\pd{}\epsilon_j}w_{k}^{(h)}(p_1,\dotsc,p_k)=
	-\oint_{{\cal B}_j}w_{k+1}^{(h)}(p_1,\dotsc,p_k,q)
	\label{eq:Vff.3}
\end{equation}

\subsection{Variation of the temperatures}
Similarly we define for $t_\infty \equiv T$ and $t_{\hat{\lambda}_i}$,
\begin{equation}
	\Omega(p)=-\d{S}_{\alpha,\alpha^\prime}=\int_{\alpha}^{\alpha^\prime} B(p,q),\quad\, \text{i.e. 
	$\delta\Omega=[\alpha,\alpha^\prime], \Lambda=1$}
	\label{eq:VT}
\end{equation}
where $\alpha,\alpha^\prime\in\{\infty,\hat{\lambda}_1,\dotsc,\hat{\lambda}_\s\}$. 
This variation produces the following modifications of parameters
\begin{equation}
	\delta_{\Omega}\epsilon_l=0\,,\quad\delta_{\Omega}t_\beta=\delta_{\alpha,\beta}-\delta_{\alpha^\prime,\beta}
	\,,\quad\delta_{\Omega}g_j^{(m)}=0\,,\quad
	\delta_{\Omega}\lambda_i=0
	\label{eq:VT.2}
\end{equation}
which can be written as $\delta_{-\d{S}_{\alpha,\alpha^\prime}}=
\frac{\pd{}}{\pd{}t_\alpha}-\frac{\pd{}}{\pd{}t_\alpha^\prime}$. This makes sense since $\sum_\alpha t_\alpha =0$.
Again Theorem 5.1 in \cite{EOFg} enables us to write
\begin{equation}
	\left(\frac{\pd{}}{\pd{}t_\alpha}-\frac{\pd{}}{\pd{}t_\alpha^\prime}\right)w_{k}^{(h)}(p_1,\dotsc,p_k)=
	\int_{\alpha}^{\alpha^\prime}w_{k+1}^{(h)}(p_1,\dotsc,p_k,q)
	\label{eq:VT.3}
\end{equation}

\subsection{Variation of the potentials}\label{sSec:VP}

Observe that if we don't consider variations in $\cc{i,i+1}$ we have
\begin{equation}
\begin{split}
\cc{1,2}(\delta x_2 . dx_1 - \delta x_1 . dx_2) 
& = \delta x_2 . (V''_2(x_2) dx_2 - \cc{2,3}dx_3) - ((\delta V'_2)(x_2). dx_2 + V''_2(x_2) \delta x_2 .dx_2 - \cc{2,3}\delta x_3.dx_2) \\
& = -(\delta V'_2)(x_2). dx_2 +  \cc{2,3}(\delta x_3 . dx_2 - \delta x_2 . dx_3)  \\
& \vdots \\
& = -(\delta V'_2)(x_2). dx_2-\dots-(\delta V'_n)(x_n). dx_n +  \cc{n,n+1}(\delta x_{n+1} . dx_n - \delta x_n . dx_{n+1})  \\
\end{split}
\end{equation}
In particular, if the $\lambda_i$ are kept fixed, the last term $\delta x_{n+1} . dx_n - \delta x_n . dx_{n+1}$ has no pole at $\hat\lambda_i$.

\subsubsection*{\it Variations of \tops{$V_1$}{V1}}

If we vary only $V_1$, more precisely if we vary only $g_j^{(1)}$, we see that $\Omega=\cc{1,2} (\delta x_1 . dx_2 - \delta x_2 . dx_1)$ has no pole at the $\hat\lambda_i$'s, and near $\infty$, if we work at fixed $x_1$, we have $\cc{1,2}\delta x_2 \sim  \delta g_{j}^{(1)}\, x_1^{j-1}  + O(x_1^{-2})$, and in addition, since we don't vary the filling fractions, we know that $\oint_{\A_i} \Omega =0$. All these considerations imply that the variations of $V_1(x)$ are given by the same formulas as in \cite{ChEyOr}:
\begin{equation}
	\Omega(p)=-B_{\infty,j}(p) =\frac{1}{j}\Res_{q\to\infty} B(p,q)x_1^j(q)
	\label{eq:VP1}
\end{equation}
thus $\delta\Omega$ is a small circle around $\infty$ and $\Lambda(q)=\frac{1}{2i\pi}\frac{x_1(q)^j}{j}$.
With this variation it is easy to check that
\begin{equation}
	\delta_{\Omega}\epsilon_l=0\,,\quad\delta_{\Omega}t_\alpha=0
	\,,\quad\delta_{\Omega}g_m^{(k)}=\delta_{k,1}\delta_{j,m}\,,\quad
	\delta_{\Omega}\lambda_i=0
	\label{eq:VP.2}
\end{equation}
and so we can say that $\delta_{-B_{\infty,j}}=\frac{\pd{}}{\pd{}g_j^{(1)}}$, and from Theorem 5.1 in \cite{EOFg}:
\begin{equation}
	\frac{\pd{}}{\pd{}g_j^{(1)}}w_{k}^{(h)}(p_1,\dotsc,p_k)=
	\Res_{\infty}\frac{x_1(q)^j}{j} w_{k+1}^{(h)}(p_1,\dotsc,p_k,q).
	\label{eq:VP.3}
\end{equation}

\subsubsection*{\it Variations of \tops{$V_2,\dotsc,V_n$}{V2,...,Vn}}

For the other potentials $V_k$ with $2\leq k\leq n$, if we vary $g_j^{(k)}$, near $\infty$, at fixed $x_1$, we have $\delta x_2 = \cc{1,2}\delta (V'_1(x_1)-T/x_1+O(x_1^{-2}) ) = O(x_1^{-2})$, therefore $\Omega$ has no pole at $\infty$.
We have seen that the pole of $\Omega$ at $\hat\lambda_i$ is given by $\delta(V'_k)(x_k) dx_k$, therefore near $\hat\lambda_i$ we have $\Omega \sim - x_k^{j-1} dx_k$.
This implies that
\begin{equation}
	\Omega(p)= -\sum_i B_{\hat\lambda_i,k,j}(p)= -\frac{1}{j} \sum_i \Res_{q\to\hat\lambda_i} B(p,q)x_k^j(q)
	\label{eq:VPk}
\end{equation}
thus $\pd{}\Omega$ is a contour which surrounds all $\hat\lambda_i$ (and no other poles), and $\Lambda(q)=-\frac{1}{2i\pi}\frac{x_k(q)^j}{j}$.
Then, we can say that $\delta_{-\sum_i B_{\hat\lambda_i,k,j}}=\frac{\pd{}}{\pd{}g_j^{(k)}}$, and from Theorem 5.1 in \cite{EOFg}:
\begin{equation}
	\frac{\pd{}}{\pd{}g_j^{(k)}} w_{l}^{(h)}(p_1,\dotsc,p_l)=
	\sum_i \Res_{\hat\lambda_i}\frac{x_k(q)^j}{j} w_{l+1}^{(h)}(p_1,\dotsc,p_l,q).
	\label{eq:VP.3k}
\end{equation}

\subsection{Variation of the \tops{$\lambda_i$}{L}'s}\label{sSec:Vlambdai}

Similarly, we see that when we vary $\lambda_i$, $\Omega$ has no pole at $\infty$, and near $\hat\lambda_i$, it behaves like $-dx_n$.
Therefore we have:
\begin{equation}
	\Omega(p)= T\frac{l_i}{N}\,\, B_{\hat\lambda_i}(p) = \Res_{q\to\hat\lambda_i} B(p,q) x_n(q) = 
	T\frac{l_i}{ N}\,\, \frac{B(p,\hat\lambda_i)}{dx_{n+1}(\hat\lambda_i)}
	\label{eq:Vlambdai}
\end{equation}

\begin{equation}
	\frac{\pd{}}{\pd{}\lambda_i} w_{l}^{(h)}(p_1,\dotsc,p_l)=
	 T\frac{l_i}{ N}\,\,\Res_{\hat\lambda_i} w_{l+1}^{(h)}(p_1,\dotsc,p_l,q)\, x_n(q)\, .
	\label{eq:varlambdai}
\end{equation}

\subsection{Variation of the \tops{$\cc{k,k+1}$}{g(i,i+1)}}\label{sSec:g'sint}

Again, allowing variations in $\cc{k,k+1}$ we find that $\Omega$ behaves like $x_k\d{}x_{k+1}$, therefore
\begin{equation}
	\Omega(p)=\sum_{i}B_{\hat{\lambda}_i,k\to k+1}=\sum_i \Res_{\hat{\lambda_i}} B(p,q) x_k(q) x_{k+1}(q)
	\label{eq:g's}
\end{equation}

\begin{equation}
	\frac{\pd{}}{\pd{} \cc{k,k+1}} w_{l}^{(h)}(p_1,\dotsc,p_l)=
	 \sum_i \,\,\Res_{\hat\lambda_i} w_{l+1}^{(h)}(p_1,\dotsc,p_l,q)\, x_k(q) x_{k+1}(q)\, .
	\label{eq:VP.varci}
\end{equation}

\subsection{Summary of moduli}

Using Cauchy formula, we may write:
\begin{equation}
\cc{1,2}x_2(p) dx_1(p) = - \Res_{q\to p} dS_{q,o}(p)\, \cc{1,2}x_2(q) dx_1(q) 
\end{equation}
Then, we move the integration contour, and we take into account the boundary terms using Riemann bilinear identity, we get:
\begin{equation}
\cc{1,2}x_2(p) dx_1(p) =  \Res_{q\to \infty,\hat\lambda_i} dS_{q,o}(p)\, \cc{1,2}x_2(q) dx_1(q)  
+ 2i\pi \sum_i \epsilon_i du_i(p)
\end{equation}
The residues near the poles $\infty$ and near the $\hat\lambda_i$ are computed by the local behaviors.

\medskip
$\bullet$ near $\infty$, we have $x_2\sim V'_1(x_1) - \frac{T}{x_1} + O(x_1^{-2})$, i.e.
\begin{equation}
\begin{split}
\Res_{q\to \infty} dS_{q,o}(p)\, \cc{1,2}x_2(q) dx_1(q)
&= \Res_{q\to \infty} dS_{q,o}(p)\, d V_1(x_1(q)) - T \Res_{q\to \infty} dS_{q,o}(p)\, \frac{dx_1(q)}{x_1(q)} \\
&= - \Res_{q\to \infty} B(q,p)\, V_1(x_1(q)) + T dS_{\infty,o}(p) \\
&= - \sum_j \frac{g_j^{(1)}}{j} \Res_{q\to \infty} B(q,p)\, (x_1(q))^j + T dS_{\infty,o}(p) \\
&= \sum_j \frac{g_j^{(1)}}{j} B_{\infty,j}(p) + T dS_{\infty,o}(p) \\
\end{split}
\end{equation}

\medskip
$\bullet$ near $\hat\lambda_i$, we have 
\begin{equation}
\begin{split}
\cc{1,2}x_2 dx_1 
&= \cc{1,2}(d(x_1 x_2) - x_1 dx_2) 
= \cc{1,2}d(x_1 x_2) - V'_2(x_2) dx_2 + \cc{2,3}x_3 dx_2 \\
& \qquad \quad \vdots  \\
&= d(\cc{1,2}x_1 x_2 - V_2(x_2) + \cc{2,3}x_2 x_3 - V_3(x_3)  \dots \\
&\hspace{40pt}+ \cc{n-1,n}x_{n-1} x_n - V_n(x_n) 
+ \cc{n,n+1}x_{n} x_{n+1}) - \cc{n,n+1}x_n dx_{n+1} \\
&\sim \,\, d(\cc{1,2}x_1 x_2 - V_2(x_2) + \cc{2,3}x_2 x_3 - V_3(x_3)  \dots \\
&\hspace{40pt}+ \cc{n-1,n}x_{n-1} x_n 
- V_n(x_n) + \cc{n,n+1}x_{n} x_{n+1}) 
- T\frac{l_i}{N}\, \frac{dx_{n+1}}{x_{n+1}-\lambda_i} + \O(1) \\
\end{split}
\end{equation}
Therefore we have:
\begin{equation}
\begin{split}
\Res_{q\to \hat\lambda_i} &dS_{q,o}(p)\, \cc{1,2}x_2(q) dx_1(q)=\\
&= -T\frac{l_i}{N}\, dS_{\hat\lambda_i,o}(p) -
\Res_{q\to \hat\lambda_i} dS_{q,o}(p)\, d(V_2+\dots+V_{n} - \cc{1,2}x_1x_2 -\dots -\cc{n,n+1}x_{n} x_{n+1}) \\
&= -T\frac{l_i}{N}\, dS_{\hat\lambda_i,o}(p) 
+ \Res_{q\to \hat\lambda_i} B(q,p)\, (V_2+\dots+V_{n} - \cc{1,2}x_1x_2 -\dots -\cc{n,n+1}x_{n} x_{n+1}) \\
&= -T\frac{l_i}{N}\, dS_{\hat\lambda_i,o}(p) 
+ \sum_{k=2}^n \sum_j \frac{g_j^{(k)}}{j}\, \Res_{q\to \hat\lambda_i} B(q,p)\, (x_k(q))^j 
- \sum_{k=1}^n   \cc{k,k+1}\Res_{q\to \hat\lambda_i} B(q,p) x_k(q) x_{k+1}(q) \\
\end{split}
\end{equation}

All this can be summarized as:
\begin{equation}
\begin{split}
\cc{1,2} x_2 dx_1 
&= 2i\pi \sum_i \epsilon_i du_i + \sum_i \frac{T l_i}{N}\, dS_{\infty,\hat\lambda_i} + \sum_j g_{j}^{(1)} B_{\infty,j}  + \sum_{k=2}^n \sum_j g_j^{(k)} \sum_i B_{\hat\lambda_i,k,j}  \\
& + \sum_i \frac{T l_i}{N} \lambda_i B_{\hat\lambda_i} + \sum_{k=1}^{n-1} \cc{k,k+1}\,  \sum_i B_{\hat{\lambda}_i,k\to k+1}
\end{split}
\end{equation}

Notice that $\cc{n,n+1}$ does not appear, in fact the term that would logically give the associated contribution, it is better used to encode the variations of $\lambda_i$. It is clear that the $\lambda_i$ contain already the information of $\cc{n,n+1}$.

\subsection{Topological expansion of the free energy}\label{Sec:TEFE}

With all that information we are now ready to derive the free energy ${\cal F}^{(g)}$.
The free energy $\ln{Z} = {\cal F}= \sum_g (N/T)^{2-2g} {\cal F}^{(g)}$, is determined by its derivatives:
\begin{equation}
	\frac{1}{j}\,<{\rm Tr}\, (M_k)^j >=-\frac{\pd{}{\cal F}}{\pd{}g_j^{(k)} }.
\end{equation}
The result that we wish to prove is that:
\begin{equation}
{\cal F}^{(g)} = F_g(\hat{E}^{(0)}) 
\end{equation}
where $F_g$'s are the symplectic invariants of \cite{EOFg}, for the spectral curve $\hat{E}^{(0)}$.
In particular for $g\geq 2$ we have:
\begin{equation}
F_g(\hat{E}^{(0)})  = \frac{1}{2-2g}\,\sum_\alpha \Res_\alpha w_1^{(g)}\, \Phi
\qquad , \quad d\Phi= \cc{1,2} x_2 dx_1
\end{equation}
The expressions of $F_0$ and $F_1$ are a little bit more difficult to write \cite{Kri, KoKo}, and we refer the reader to \cite{EOFg}. Notice that when there is no external field, i.e. $\Lambda=0$, ${\cal F}^{(0)}$ was already computed in \cite{eynchainloopeq}, and it coincides with $F_0$.

\medskip

The $F_g$'s of \cite{EOFg} have the property, that under any variation $\Omega$, we have:
\begin{equation}
\delta F_g = \int_{\pd{} \Omega} w_1^{(g)}(q)\, \Lambda(q)
\end{equation}
In particular with $k=1$, it proves that
\begin{equation}
\frac{\pd{} F_g}{\pd{} g_j^{(1)}} = \frac{1}{j}\, \Res_\infty w_1^{(g)}\, x_1^j = \frac{\pd{}{\cal F}^{(g)}}{\pd{}g_j^{(1)} }
\end{equation}

\medskip

Then, we prove it by recursion on the length of the chain $n$.

The $n=1$ case was done in \cite{EOFg}.
Now, assume that it is true for $n-1$.

We have just seen that ${\cal F}^{(g)}-F_g$ is independent of $V_1$, therefore we may compute it for 
the case where $V_1$ is quadratic.
When $V_1$ is quadratic, the integral over the first matrix of the chain, $M_1$, is a gaussian integral, and $M_1$ can be integrated out, so that when $V_1$ is quadratic we are left with a chain of $n-1$ matrices, 
and we get ${\cal F}^{(g)}_{n}={\cal F}^{(g)}_{n-1}$.
From the recursion hypothesis, we have ${\cal F}^{(g)}_{n-1} = F_g(\hat{E}^{(0)}(x_2,x_3))$,
and one should notice that the $F_g$'s of \cite{EOFg} have the symplectic invariance property, i.e. they are unchanged if we make a symplectic transformation of the spectral curve, or in other words, if we add an exact differential to $\cc{1,2}x_2 dx_1$. In particular we may work with $\cc{2,3}x_3 dx_2$, and thus $F_g(\hat{E}^{(0)}(x_1,x_2))=F_g(\hat{E}^{(0)}(x_2,x_3))$.
This proves the result.

\section{Other Considerations}\label{Sec:OC}

In the previous two sections, we have solved the loop equations to all orders, and we have found that the solution is given by the symplectic invariants introduced in \cite{EOFg}, for the spectral curve $\hat{E}^{(0)}(x_1,x_2)$.
As a consequence, all the properties studied in \cite{EOFg} apply.

\subsection{Symplectic transformations}

Remember that the spectral curve $\hat{E}(x_1,x_2)=0$ is equivalently given by the data of two meromorphic functions $x_1(p),x_2(p)$ on $\cal L$.
Indeed, given two meromorphic functions, it is always possible to find a polynomial relationship between them.
We shall write the spectral curve:
\begin{equation}
\hat{E}_{1,2} = \{(x_1(p),x_2(p))\,\, / \,\, p\in {\cal L}\} = (x_1,x_2)
\end{equation}
Since any $x_i$ is a meromorphic function, we can also define the following algebraic spectral curves:
\begin{equation}
\hat{E}_{i,j} = \{(x_i(p),x_j(p))\,\, / \,\, p\in {\cal L}\} = (x_i,x_j)
\end{equation}

It was found in \cite{EOFg, EOsym}, that the $F_g$'s are unchanged under symplectic transformations of the spectral curve, for instance if we add to $x_1$ any rational function of $x_2$, or if we exchange $x_1 \leftrightarrow x_2$, or if we change $x_1\to -x_1$.

For instance we could change $\cc{1,2}x_1\to \cc{1,2}x_1-V'_2(x_2) = -\cc{2,3}x_3$, and then $x_3\to -x_3$, and then recursively $\cc{i,i+1}x_i\to \cc{i,i+1}x_i-V'_{i+1}(x_i)=-\cc{i+1,i+2}x_{i+2}$. This shows that:
\begin{equation}
F_g = F_g(\hat{E}_{i,i+1}) = F_g(\hat{E}_{i+1,i})
\qquad \forall \, 1\leq i\leq n
\end{equation}

However, one should keep in mind that the correlation functions are not conserved under symplectic transformations, only the $F_g$'s are.

\subsection{Double scaling limits}\label{Sec:DSL}

We have seen that as long as the spectral curve is regular (all branch-points are simple), the $F_g$'s and all correlation functions can be computed, and it was found in \cite{EOFg} that they diverge when the curve becomes singular. 
This type singularities were already found in the one and two matrix model, and in \cite{EOFg} for generic spectral curves, but it is still important to show that it appears in this context too. 

It was found in \cite{EOFg}, that if the spectral curve depends on some coupling constant ($T$ for instance), if the spectral curve develops a cusp singularity at say $T=T_c$ of the form
\begin{equation}
y\sim x^{p/q}
\end{equation}
then the $F_g$'s diverge as
\begin{equation}
F_g \sim (1-\frac{T}{T_c})^{(2-2g)\frac{p+q}{p+q-1}}\,\, \tilde{F}_g
\end{equation}
where $\tilde{F}_g=F_g(\tilde{E})$ are the symplectic invariants of another spectral curve $\tilde{E}$ which is the blow up of the vicinity of the singularity, and which is the spectral curve of the $(p,q)$ minimal model \cite{DKK, KazakovGQ}. All this is detailed in \cite{EOFg} and we refer the reader to that article for more details.

\medskip

As usual, singularities of formal series are related to the large order asymptotic expansion of the general term of the series \cite{ZJDFG}, and the double scaling limit is thus related to the asymptotic enumeration of large discrete surfaces, and in some sense to their continuous limit, i.e. Riemann surfaces.
A $(p,q)$ minimal model may occur as soon as two of the $V_i$'s have degree larger than p and q \cite{DKK}.
Here, we see that the double scaling of the chain of matrices describes a $(p,q)$ minimal model on a random lattice. This is related to the Liouville conformal field theory coupled to minimal models $(p,q)$. This phenomenon is expected \cite{KazakovGQbis, ZJDFG, KPZ, DKK} and is already known to be present in the one and two matrix models and more generaly in \cite{EOFg}.

\subsection{Modular transformations and holomorphic anomaly equations}\label{Sec:MT}

In order to compute the $F_g$'s and the solution of loop equations, we have made a choice of cycles $\A_i$, related to the choice of the minimum around which the formal matrix integral is defined.
However, it is interesting to see what happens if one makes a different choice of cycles, i.e. if one makes a modular transformation.
This was studied in \cite{EOFg} and \cite{EOMham}.

A modular transformation changes the Bergman kernel $B(p,q)$ with a constant symmetric matrix $\kappa$:
\begin{equation}
B(p,q) \to  B(p,q) + 2i\pi\,\sum_{i,j} \kappa_{i,j}\, du_i(p)du_j(q)
\end{equation}
where $du_i$ are the holomorphic forms on ${\cal L}$ such that $\oint_{\A_j} du_i=\delta_{i,j}$.

In particular, if
\begin{equation}
\kappa = \frac{i}{2}\,(\Im \tau)^{-1}
\end{equation}
(where $\tau_{i,j}=\oint_{\B_j}du_i$ is the Riemann matrix of periods of ${\cal L}$) then the Bergman kernel is called Schiffer kernel and is modular invariant.

More generally, the modular transformations were computed in \cite{EOFg}, and they satisfy the so-called holomorphic anomaly equations, and that gives a strong support to the Dijkgraaf-Vafa conjecture that matrix models are topological type B string theory partition functions \cite{BCOV, EOMham}.

\subsection{Convergent matrix integrals and filling fractions}

So far, we have considered formal matrix integrals, defined by expanding the integrand in the matrix integral, near a given extrema specified by a set of filling fractions.
We worked at fixed filling fractions.

On the other hand, convergent matrix integrals should correspond to integrals over $(H_N)^n$.
The integration path can always be written as a linear combination of steepest descent paths (those used for formal integrals), and the full convergent matrix integral is obtained as a linear combination of formal matrix integrals. More precisely, the convergent matrix integral should be a sum over filling fractions of the formal ones.

The summation over filling fractions was computed in \cite{Eynff}, and just amounts to multiplication of the formal matrix integrals by a theta function. We refer the reader to \cite{BDE, Eynff} for more details.

\section{Limit of a continuous chain of matrices}\label{Sec:QMM}

In this section, we briefly explore some consequences of our method for the continuous chain of matrices.

The "matrix-model quantum mechanics", is obtained \cite{eynchain} as the limit $n\to\infty$, and with the choice:
$c_{i,i+1}=\frac{1}{\epsilon}$, and:
\begin{equation}
c_{i,i+1}=\frac{1}{\epsilon}
\quad , \qquad
V_1(x) = \epsilon{\cal V}(x,\epsilon) + \frac{x_1^2}{2\epsilon}
\quad , \qquad
V_i(x) = \epsilon{\cal V}(x,\epsilon i) + \frac{x_i^2}{\epsilon}
\end{equation}
The index $i$ is rescaled as a continuous time $t=\epsilon i$:
\begin{equation}
t = \epsilon i
\quad , \qquad
0\leq t \leq t_f= \epsilon n
\end{equation}

The matrix integral thus becomes:
\begin{equation}
Z = \int D[M(t)]\,\,\, \e{-\frac{N}{T} \int_{0}^{t_f}\, dt \,\,\,{\rm Tr} \left[  {\cal V}(M(t),t) + \frac{1}{2} \dot{M}(t)^2 \right]}
\end{equation}
The spectral curve is determined by the equations $V'_i(x_i) = c_{i,i+1} x_{i+1}+c_{i,i-1}x_{i-1}$ which become Newton's equation of motion \cite{eynchain} to leading order in $\epsilon$:
\begin{equation}
{\cal V}'(x,t) = \ddot{x}(t)
\end{equation}
and the resolvent of the first matrix is:
\begin{equation}
W(x,0) = V'_1(x_1) - c_{1,2} x_2 \sim - \dot{x}(0)
\end{equation}
The topological expansion is thus:
\begin{equation}
Z = \e{\sum_g (N/T)^{2-2g} F_g}
\end{equation}
\begin{equation}
F_g = F_g({\cal E}(t))
\quad , \qquad
{\cal E}(t)=(x(t),-\dot{x}(t))
\end{equation}
The spectral curve ${\cal E}(t)=(x(t),-\dot{x}(t))$ is thus the dispersion relation, i.e. the relationship between velocity and position, it may depend on the time $t$, but from symplectic invariance, we see that $F_g({\cal E}(t))$ is a conserved quantity, independent of the time $t$.

For example, if the potential ${\cal V}(x,t)={\cal V}(x)$ is independent of $t$, the kinetic energy $K$ is conserved and the dispersion relation is:
\begin{equation}
 \frac{1}{2}\dot{x}^2 - {\cal V}(x(t)) = K
\end{equation}
and the spectral curve is:
\begin{equation}
{\cal E}(t)=(x(t), \sqrt{2({\cal V}(x(t)) + K)})
\end{equation}

Consequences of those relations need to be further explored, and we leave the continuous chain of matrices for another work.

\section{Conclusion}\label{Sec:Concl}

We have computed explicitly the topological expansion of the chain of matrices with an external field, and we have found that the $F_g$'s are precisely the symplectic invariants of \cite{EOFg}.

We have also computed some of the correlation functions, but not all of them, in particular we have not computed mixed traces (which count discrete surfaces with non-trivial boundary conditions).
Mixed traces were computed in the 2-matrix model case in \cite{EOsym, EOmixedplanar}, and it would be interesting to see how that could be extended to the chain of matrices.

We have also briefly started to explore the limit of matrix quantum mechanics, i.e. the limit of an infinite chain of matrices, but this topic needs to be studied in deeper details.

\section*{Acknowledgments}
We would like to thank L. Cantini and N. Orantin, J.B. Zuber, for useful and fruitful discussions on this subject.
This work is partly supported by the Enigma European network MRT-CT-2004-5652, by the ANR project G\'eom\'etrie et int\'egrabilit\'e en physique math\'ematique ANR-05-BLAN-0029-01, by the Enrage European network MRTN-CT-2004-005616,
by the European Science Foundation through the Misgam program,
by the French and Japaneese governments through PAI Sakurav, by the Quebec government with the FQRNT.


\begin{thebibliography}{99}



\bibitem{ambjornrmt} J. Ambj{\o}rn, B. Durhuus, J. Fr{\"o}hlich, "Diseases of triangulated random surface models, and possible cures", Nuclear Physics B, Volume 257, p. 433-449.

\bibitem{BergSchif} S. Bergman, M. Schiffer, ``Kernel functions and elliptic differential equations in mathematical physics'', Academic Press Inc., Publishers, New York, NY, 1953.

\bibitem{BCOV} M.Bershadsky, S.Cecotti, H.Ooguri and C.Vafa,
``Kodaira-Spencer theory of gravity and exact results for quantum string amplitudes'',
{\em Commun. Math. Phys.} {\bf 165} (1994) 311.

\bibitem{Marco2} M. Bertola, ''Two-matrix model with semiclassical potentials and extended Whitham hierarchy'',
{\em  J.Phys.} {\bf A39} 8823-8856  (2006),  hep-th/0511295.

\bibitem{MarcoF} M. Bertola, ''Free Energy of the Two-Matrix
Model/dToda Tau-Function'',
preprint CRM-2921 (2003), hep-th/0306184.

\bibitem{BDE}
G. Bonnet, F. David, B. Eynard,
Breakdown of universality in multi-cut matrix models,
J.Phys. A: Math. Gen. {\bf 33} (2000) 6739-6768.

\bibitem{BIPZ} E. Brezin, C. Itzykson, G. Parisi, and J.B. Zuber,
{\em Comm. Math. Phys.} {\bf 59}, 35  (1978).

\bibitem{eynch} L. Chekhov, B. Eynard, Hermitean matrix model free energy: Feynman graph technique for all genera, JHEP 009 P0206/5, hep-th/0504116.

\bibitem{ChEyOr} L.~Chekhov, B.~Eynard, N.~Orantin, Free energy topological expansion for the 2-matrix model, JHEP 0612 (2006) 053, math-ph/0603003.

\bibitem{davidRMT} F. David, "Planar diagrams, two-dimensional lattice gravity and surface models",
Nuclear Physics B, Volume 257, p. 45-58.

\bibitem{DKK} J.M.Daul, V.Kazakov, I.Kostov,
``Rational Theories of 2D Gravity from the Two-Matrix Model'',
{\em Nucl.Phys.} {\bf B409} (1993) 311-338, hep-th/9303093.

\bibitem{DW} R.Dijkgraaf and E.Witten,
``Mean field theory, topological field theory, and multimatrix models'',
{\em Nucl.Phys.} {\bf B342} (1990) 486--522.

\bibitem{ZJDFG} P. Di Francesco, P. Ginsparg, J. Zinn-Justin,
``2D Gravity and Random Matrices'',
{\em Phys. Rep.} {\bf 254}, 1 (1995).


\bibitem{eynchain} B. Eynard, Correlation functions of eigenvalues of multi-matrix models, and the limit of a time dependent matrix, Journal of Physics A 40 (1998) 8081, cond-mat/9801075.

\bibitem{eynchainloopeq} B. Eynard, Master loop equations, free energy and correlations for the chain of matrices, JHEP11(2003)018, hep-th/0309036.

\bibitem{eyn1loop} B. Eynard, Topological expansion for the 1-hermitian matrix model correlation functions,\\ JHEP/024A/0904, hep-th/0407261.

\bibitem{EOmixedplanar} B. Eynard, N. Orantin, Mixed Correlation Functions in the 2-Matrix Model, and Bethe Ansatz, JHEP/0508 (2005) 028, hep-th/0504029.

\bibitem{eoloop2mat} B. Eynard, N. Orantin, Topological expansion of the 2-matrix model correlation functions: diagrammatic rules for a residue formula, math-ph/0504058, JHEP 12(2005) 034.

\bibitem{EOFg} B. Eynard, N. Orantin, Invariants of algebraic curves and topological expansion, math-ph/0702045, Communications in Number Theory and Physics, Vol 1, Number 2, p347-452.
	
\bibitem{EOsym} B. Eynard, N. Orantin, Topological expansion of mixed correlations in the hermitian 2 Matrix Model and x-y symmetry of the $F_g$ algebraic invariants,  arXiv:0705.0958 [math-ph], to appear in J.Phys A.

	 
 \bibitem{Eynff} B. Eynard, Large N expansion of convergent matrix integrals, holomorphic anomalies, and background independence, 	arXiv:0802.1788v1 [math-ph].

\bibitem{EOMham} B. Eynard, M. Marino, N. Orantin, Holomorphic anomaly and matrix models, hep-th/0702110, JHEP 089P 0307.


\bibitem{eynform} B.Eynard,
`` Formal matrix integrals and combinatorics of maps'', \\
math-ph/0611087.

\bibitem{eynMgnkappa}B. Eynard, Recursion between Mumford volumes of moduli spaces, arXiv:0706.4403v1 [math.AG].

\bibitem{Farkas} H.M. Farkas, I. Kra, ''Riemann surfaces'' 2nd edition, Springer Verlag, 1992.

\bibitem{Fay} J.D. Fay, ''Theta functions on Riemann surfaces'', Springer Verlag, 1973.

\bibitem{Kazakovloop} V.A. Kazakov,
"The appearance of matter fields from quantum fluctuations of 2D-gravity", Modern Physics Letters A, Vol. 4, No. 22 (1989) 2125-2139.

\bibitem{Kazakov} V.A. Kazakov, ``Ising model on a dynamical
planar random lattice: exact solution'',
{\em Phys Lett.} {\bf A119}, 140-144 (1986).

\bibitem{KazakovGQ} V.A. Kazakov , "Bilocal regularization of models of random surfaces", Phys.Lett.B150:282-284,1985.

\bibitem{KazakovGQbis} V.A. Kazakov, A. A. Migdal, I.K. Kostov, "Critical Properties Of Randomly Triangulated Planar Random Surfaces", Phys.Lett.B157:295-300,1985.

\bibitem{KoKo} A.Kokotov, D.Korotkin,
`` Bergman tau-function on Hurwitz spaces and its applications'',
math-ph/0310008.
\bibitem{KPZ} V.G. Knizhnik, A.M. Polyakov, A.B. Zamolodchikov, Mod. Phys. Lett {\bf A3} (1988) 819.

\bibitem{kontsevitch} M. Kontsevich,
``Intersection theory on the moduli space of curves and the matrix Airy function'',
{\em Funk. Anal. Prilozh.} {\bf 25} (1991) 50-57; Comm. Math. Phys. {\bf 147} (1992), no 1. 1-23;
Max-Planck Institut preprint MPI/91-47, MPI/91-77.

\bibitem{Kri} I.Krichever
``The $\tau$-function of the universal Whitham
hierarchy, matrix models and topological field theories'',
{\em Commun.Pure Appl.Math.} {\bf47} (1992) 437; hep-th/9205110



\bibitem{thooft} G. 't Hooft, {\em Nuc. Phys.} {\bf B72}, 461 (1974).





\end{thebibliography}
\end{document}